\documentclass[aps,showpacs]{revtex4}
\usepackage{amsmath}
\usepackage{amssymb}
\usepackage{graphicx}
\usepackage{dcolumn}
\usepackage{bm}
\usepackage{epstopdf}

\newcommand{\la}{\lambda}

\newcommand{\om}{\omega}

\newcommand{\prt}{\partial}

\begin{document}

\title{
Transcritical flow of a Bose-Einstein condensate through a penetrable barrier}

\author{A.M. Leszczyszyn$^{1}$}
\email{A.M.Leszczyszyn@lboro.ac.uk}
\author{G.A. El$^{1}$}
\email{G.El@lboro.ac.uk}
\author{Yu.G. Gladush$^{2}$}
\email{gladush@isan.troitsk.ru}
\author{A.M. Kamchatnov$^{2}$}
\email{kamch@isan.troitsk.ru}

\affiliation{
$^1$Department of Mathematical Sciences, Loughborough
University, Loughborough LE11 3TU, UK \\
$^2$Institute of Spectroscopy, Russian Academy of Sciences, Troitsk,
Moscow Region, 142190, Russia\\ }

\begin{abstract}
The problem of the transcritical flow of a Bose-Einstein condensate through a wide repulsive
penetrable barrier is studied analytically using the combination of the localized
``hydraulic''  solution of the 1D Gross-Pitaevskii equation
and the solutions of the Whitham modulation equations describing the resolution of
the upstream and downstream discontinuities through dispersive shocks.
It is shown that within the physically reasonable range of parameters the downstream
dispersive shock is attached to the barrier and effectively
represents the train of very slow dark solitons, which can be observed in experiments.
The rate of the soliton emission, the amplitudes of the solitons
in the train and the drag force are determined in terms of the BEC oncoming flow
velocity and the strength of the potential barrier. A good agreement with direct numerical solutions is demonstrated. Connection with recent experiments is discussed.
\end{abstract}


\maketitle

\section{Introduction}

The problem of the fluid flow past an obstacle is of great importance for both
normal fluids and superfluids. In dynamics of viscous compressible fluids it
is closely related to the theory of viscous shocks (see, e.g., \cite{CF48,LL6}).
In the theory of surface water waves it has lead to the development of a detailed
description of ship waves and corresponding drag forces (see, e.g., \cite{lamb}).
In superfluid dynamics,  understanding of the nature of the critical velocity
in the flow past an obstacle was crucially important for the development of the
superfluidity theory \cite{landau,onsager,feynman}. Therefore it is natural that
this problem has attracted much interest in the context of dilute gaseous
Bose-Einstein condensates (BECs) dynamics where formation of dispersive shocks
and ``ship waves'' has been studied both theoretically
\cite{damski04,kgk04,ek06,egk06,hoefer,caruso,gegk07,kp08,gsk08}
and experimentally \cite{hoefer,caruso,simula}.
So far the main attention was paid to physically multi-dimensional problems
which in some cases could be asymptotically reduced to effectively one-dimensional (1D) models
for the purposes of mathematical convenience. However, there exist essentially
1D situations when the phenomena observed are specific to the 1D case only
(see, for instance, the discussion in \cite{ap04}). One of such prototypical
problems arises in the description of
the water flow over a bottom ridge \cite{akylas84, baines84, gs86}. In the BEC
context, a similar  problem arises in the study of the BEC flow induced by
a penetrable barrier moving through an elongated BEC, as it has been
recently studied in the experiment \cite{ea07}. The most characteristic
phenomenon observed in this experiment is that a chain of  dark solitons is
generated for a finite interval, $v_-<v<v_+$, of the barrier velocities $v$.
The existence of certain  threshold velocity $v_-$ agrees with the main concepts of
the superfluidity theory
introduced by Landau \cite{landau} and with earlier experiments
\cite{raman} on the appearance of a critical velocity for an ``impenetrable'' obstacle
moving through the condensate at different velocities. Taking into account the fact that
formation of vortices is impossible in a quasi-one-dimensional setting, one can
easily deduce from the dispersion relation for linear Bogoliubov excitations
in weakly non-ideal Bose gas that the critical velocity $v_-$ coincides with the
sound velocity, and this conclusion was confirmed by the experiment \cite{raman}.
However,  nonlinear effects modify considerably the Landau
criterion based on the idea of the generation of linear excitations at velocities
greater than the critical one. For example, in the case of wide smooth potentials
used in \cite{ea07} one has to take into account nonlinear effects which
make possible the generation of nonlinear excitations (dark solitons) and which reduce
the critical velocity $v_-$ to a subsonic value \cite{hakim}. It is remarkable
that this nonlinear mechanism is effective for a finite interval of the barrier velocities only.
As was first noticed in \cite {law}, there exist such special forms of the barrier
potential that no radiation is generated at supersonic velocities greater than
the second critical velocity $v_+$, and this result was confirmed numerically for
the potentials of a more general form. As was noted in \cite{hakim2}, while
generically some linear radiation still exists for $v>v_+$, its amplitude decreases exponentially
with the growth of the ratio of the obstacle size to the healing length so that ``broad and smooth'' potentials can always be considered as
``radiationless'' outside the interval $[v_-, v_+]$. These results
were also confirmed by the numerical simulations of waves generated by a supersonic motion of
a repulsive rectangular potential \cite{pavloff} and oscillatory
supersonic motion of the Gaussian potential \cite{radouani}. Thus, for smooth wide
penetrable potentials the production of solitons is effective only in
a finite range of the barrier velocities.
In the present paper, motivated by the results of the experiment \cite{ea07} and
the theoretical setting of \cite{hakim,baines84,gs86}, we develop
full asymptotic theory of the transcritical  BEC flow for the case of broad repulsive potentials.

Although the experiment \cite{ea07} was performed with a dense BEC for which the
radial motion of the gas is essential, we shall confine ourselves here to the case
of a rarefied gas with a frozen radial motion when the full Gross-Pitaevskii (GP)
equation can be reduced to the 1D nonlinear Schr\"odinger (NLS) equation
(see, e.g., \cite{perez98}). This limiting case reproduces all the main characteristic
features of the phenomenon, it admits exhaustive analytical treatment, and such a
setup can be quite suggestive for future experiments.

\section{Flow of a Bose-Einstein condensate through a penetrable barrier}

\subsection{Mathematical model}

Engels and Atherton \cite{ea07} considered a wide penetrable barrier moving with
constant speed through an elongated BEC confined to a cigar-shaped trap, as it
is illustrated schematically in Fig.~1.
\begin{figure}[bt]
\begin{center}
\includegraphics[width=6cm,clip]{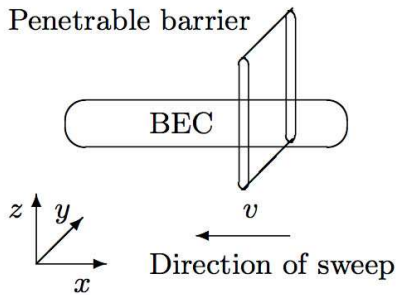}
\caption{Sketch of experimental setup in Ref.~\cite{ea07}. The potential barrier
created by the laser beam is moved through the BEC along the $-x$ direction.
The trap tightly confines the BEC in the radial $y,z$ directions.}
\end{center}\label{fig1}
\end{figure}
In the experiment, the role of the barrier was played by a repulsive potential
created by the laser
beam which is swept through the BEC with the velocity $v$ along the $-x$ direction.
If the condensate is rarefied enough, then its dynamics is described by the
forced NLS equation which in standard non-dimensional units has the form
\begin{equation}\label{3-1}
    i\psi_t+\tfrac12\psi_{xx}-|\psi|^2\psi=V(x+vt)\psi ,
\end{equation}
where $V(x+vt)>0$ is the moving potential barrier. It is convenient to
transform this equation by means of the substitution
\begin{equation}\label{3-2}
    \psi(x,t)=\sqrt{\rho(x,t)}\exp\left(i\int^x u(x',t)dx'\right)
\end{equation}
to a hydrodynamic-like form and pass to the reference frame moving with the velocity $-v$
by introducing $x'=x+vt,\,u'=u+v$:
\begin{equation}\label{3-3}
    \begin{split}
    \rho_t+(\rho u')_{x'}=0,\\
    u'_t+u'u'_{x'}+\rho_{x'}+\left(\frac{\rho_{x'}^2}{8\rho^2}
    -\frac{\rho_{x'x'}}{4\rho}\right)_{x'}+V_{x'}(x')=0.
    \end{split}
\end{equation}
Here $\rho$ and $u$ are the condensate density and velocity, respectively.
It is supposed that in this reference frame and in our non-dimensional units
the flow at infinity satisfies the conditions
\begin{equation}\label{3-4}
    \rho\to1,\quad u'\to v \quad \text{as} \quad |x|\to\infty,
\end{equation}
that is the length of the elongated condensate is assumed to be much longer than the size
of  the wave structures  generated in the flow. Thus, equations (\ref{3-3})-(\ref{3-4})
represent an idealized mathematical model for the description of
BEC dynamics in the configuration of our interest.

The study of the problem (\ref{3-3})-(\ref{3-4}) was initiated  in Ref.~\cite{hakim}
where the time-independent flows  induced by a slowly varying in space barrier
moving with subcritical velocity were analyzed.
Unsteady  supercritical flows were considered in \cite{hakim} numerically and it was
noticed there that the results are reminiscent of those for the flow of a stratified
fluid over a broad localized topography \cite{akylas84,gs86}. In \cite{gs86, s87}
the shallow-water flow past topography
problem was studied in the framework of the forced Korteweg-de Vries (KdV) equation,
and this approach provides a clue to solving our NLS equation problem
(\ref{3-3})-(\ref{3-4}). The first step in this direction is the study of the
stationary solutions of Eqs.~(\ref{3-3})-(\ref{3-4}) in the case of a wide and
smooth potential $V(x)$ so that the dispersion terms in (\ref{3-3}) can be
neglected and one can take advantage of the so-called ``hydraulic'' or
``hydrostatic'' approximation.

\subsection{Hydraulic solution}
\subsubsection{General construction}
Let the potential $V(x)$ be localized in the interval $(-l, l)$, where $l \gg 1$
(i.e. the spatial range of the potential in dimensional units is supposed to be much
greater that the healing length of the BEC). Then one can make use of the hydraulic
approximation by assuming that  the characteristic length at which all variables
change has the order of magnitude of $l$. Thus one can neglect  the terms with
the higher derivatives in
(\ref{3-3}) and consider the stationary solutions described by the system
\begin{equation}\label{4-2}
    \begin{split}
    (\rho u)_x=0,\\
    uu_x+\rho_x+V_x(x)=0,
    \end{split}
\end{equation}
where we have omitted primes for convenience of the notation. Actually, these
are equations for the stationary hydraulic flow in the reference frame with
the barrier at rest (we note that the flow direction in our paper is opposite to that
in \cite{hakim}). Both equations (\ref{4-2}) are readily  integrated once to give
\begin{equation}\label{4-3}
    \rho u=v,\quad \tfrac12u^2+\rho+V(x)=\tfrac12v^2+1 \, ,
\end{equation}
where the integration constants are found from the boundary conditions
(\ref{3-4}). Equations (\ref{4-3}) thus  define a  stationary flow that connects
smoothly to $\rho=1$, $u=v$ at both infinities.
Eliminating  $\rho$ we obtain an implicit equation for  the flow velocity $u$ as
a function of $x$,
\begin{equation}\label{4-4}
    V(x)=F(u),\quad \hbox{where} \quad F(u)=\tfrac12(v^2-u^2)-\frac{v}u+1.
\end{equation}
For (\ref{4-4}) to have a solution defined for all $x$ one should require that
\begin{equation}\label{ineq}
V_m  \le \tfrac12v^2-\tfrac32v^{2/3}+1 \, ,
\end{equation}
where $V_m= \max V(x)$. Indeed, the function $F(u)$ varies within  the interval
   $ -\infty<F(u)\leq\mu(v)$,
where $\mu(v)=\mathrm{max}\{F(u)\}=\tfrac12v^2-\tfrac32v^{2/3}+1$.
So for the equation (\ref{4-4}) to have a real solution for all $x$
the interval $[0, V_m]$  must lie within the range of the function $F(u)$
(the condition (\ref{ineq}) in the present superfluid context was  obtained in
\cite{hakim, hakim2} but can also be found in earlier
studies on shallow-water flows past topography --- see, for instance \cite{baines84}).
\begin{figure}[bt]
\begin{center}
\includegraphics[width=8cm,height=5cm,clip]{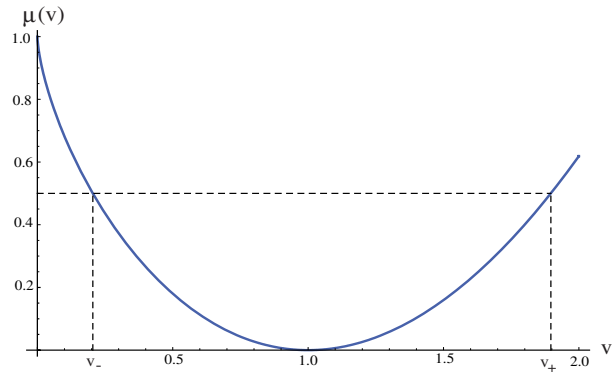}
\caption{Plot of the function $\mu(v)$; for $V_m=0.5$ the critical values are
equal to $v_- \approx 0.2,\,v_+ \approx 1.9$.}
\end{center}\label{fig2}
\end{figure}

Inequality (\ref{ineq}) defines the subcritical $v \le v_-$ and supercritical
$v \ge v_+$ regimes where $v_{\pm}$ are the roots of the equation
(see Fig.~2)
\begin{equation}\label{5-4}
    \mu(v)=\tfrac12v^2-\tfrac32v^{2/3}+1=V_m
\end{equation}
and  we have $v_-=0$ for $V_m\geq 1$.
If $V_m\ll 1$, then $v_{\pm}$ must be close to unity so one can easily find
that up to the second order in the small parameter $V_m^{1/2}$ the critical
velocities are equal to
\begin{equation}\label{5-6}
    v_{\pm} \approx1\pm\sqrt{\frac{3V_m}2}+\frac{V_m}{12}.
\end{equation}
In the {\it transcritical} regime $v_-< v < v_+$ condition (\ref{ineq}) does
not hold so it is natural
first to look closer at what happens with the hydraulic solution when $v$ approaches
its boundaries from outside.
We now look at the dependence of the flow velocity $u(x)$ and the density $\rho(x)$ on the
space coordinate $x$ determined by equations (\ref{4-3}), (\ref{4-4}) which can be re-written as
\begin{equation}\label{5-7}
    \tfrac12u^2+\frac{v}u+V(x)=\tfrac12v^2+1
\end{equation}
and
\begin{equation}\label{5-8}
    \frac{v^2}{2\rho^2}+\rho+V(x)=\tfrac12v^2+1,
\end{equation}
respectively. Formally, the solution to (\ref{5-7})  (or (\ref{5-8})) has two branches  only
one of which satisfies the necessary boundary conditions (\ref{3-4})  so
just this branch should be considered as the physical one. These two branches of the solution
for the flow velocity $u(x)$ are plotted in Fig.~3 for the case of the
potential
\begin{equation}\label{6-1}
    V(x)=\frac{V_m}{\cosh(x/\sigma)}
\end{equation}
with $V_m=0.5$, $\sigma=2$, for which Eq.~(\ref{5-4}) yields $v_- \approx 0.2 $,
$v_+ \approx 1.9 $.
We see that the subcritical ($v\leq v_-$)
and supercritical ($v\geq v_+$) hydraulic solutions have similar structure but
are characterized by the ``exchanged'' relative positions of the physical and
nonphysical branches.

\begin{figure}[bt]
\begin{center}
\includegraphics[width=8cm,height=5cm,clip]{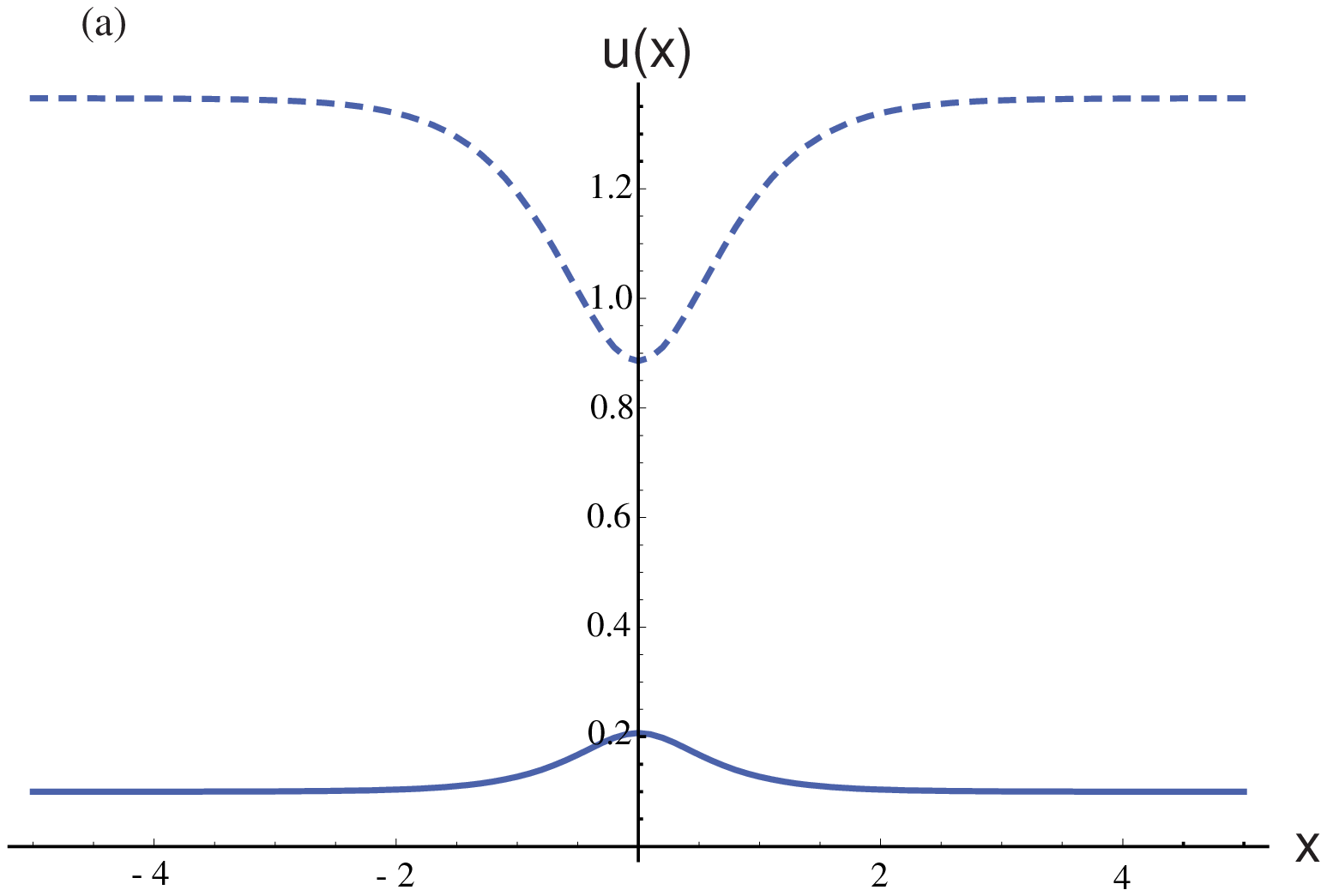}
\hspace{0.5cm}
\includegraphics[width=8cm,height=5cm,clip]{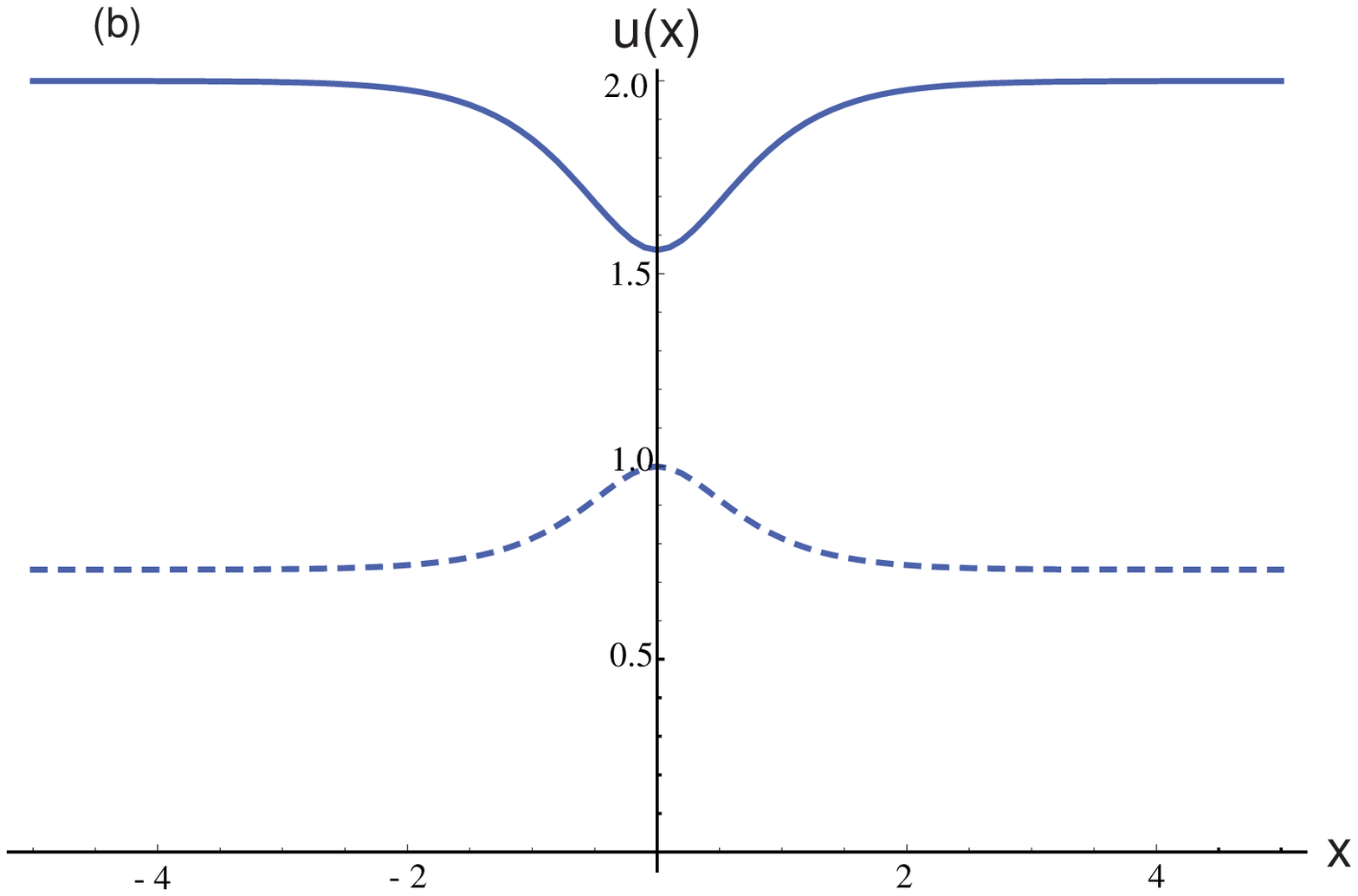}
\caption{Plots of the fluid velocity $u(x)$ in the hydraulic solution (\ref{5-7})  for the
potential barrier (\ref{6-1}) for which $v_- \approx 0.2 $, $v_+ \approx 1.9 $;
(a) corresponds to
subcritical velocity $v=0.1$ and (b) to supercritical velocity $v=2.0$. The
physical branches are shown by solid lines and nonphysical  by
dashed lines.}
\end{center}\label{fig3}
\end{figure}

In Fig.~4 we have plotted two series of hydraulic solutions for different subcritical
and supercritical values of the velocity $v$.  As one can see, the physical and
nonphysical families are separated by two separatrices.
When one approaches
$v=v_-$ from below (Fig.~4a), the physical branch becomes more and more
pointed up at the center of the barrier ($x=0$), and when  $v=v_-$, it bifurcates
into the separatrix line.   This separatrix solution satisfies the necessary
boundary condition $v=v_-$  as $x\to-\infty$ (effectively at $x=-l$)  but it fails to
satisfy the same equilibrium condition at $x \to + \infty$ so one can expect that
this bifurcation will be accompanied by
the generation of an unsteady flow downstream.

Indeed, the analysis of the  near-critical NLS flow  through the delta-functional
potential $V(x)$ in \cite{hakim}, \cite{brachet}
shows that for some $v=v_{cr}(V_m)$ the flow loses its stability  through
the saddle-node bifurcation resulting in the generation of  dark solitons
for $v>v_{cr}$. The numerical simulations in \cite{hakim} also suggest that
one can expect a qualitatively similar scenario with the soliton train generation
for slowly varying potentials moving with the supercritical velocity.

To describe this  generation of solitons in the NLS flow past broad potential barrier quantitatively, we shall take advantage of the analytical construction proposed in \cite{gs86}, where the
transcritical shallow-water flow past extended localized topography was considered in the framework of the forced KdV equation.  The key in this construction is the assumption (confirmed numerically {\it a posteriori}) of the
existence, for certain interval of the flow velocities, of the local steady {\it transcritical} transition over the forcing region,
which is described by the relevant hydraulic solution. This solution does not satisfy the equilibrium boundary conditions outside the spatial range $[-l,l]$ of the forcing, so one introduces discontinuities at $x=\pm l$, which are resolved
into the equilibrium state with the aid of unsteady nonlinear wave trains (undular bores). By applying a similar assumption here for $v=v_-$ we consider the  described above separatrix  hydraulic solution as a local one defined
on the interval $[-l, l]$. Then we get a discontinuity at $x=l$, which
should be further resolved into the undisturbed state $u=v, \rho =1$ as $x \to \infty$ via the {\it dispersive shock wave}.
Thus, when the BEC flow velocity is equal to
$v=v_-$,  the formation of the dispersive shock downstream the
potential barrier is expected.

In a similar way, when we get closer to $v=v_+$
from above (Fig.~4b), the physical solution becomes pointed down, and at $v=v_+$ it
bifurcates into the separatrix line which satisfies the boundary condition
$u=v_+$  as $x\to+\infty$ (effectively at $x=l$) and a discontinuity
$u=u_-\neq v_+$ occurs at the left edge $x=-l$. Hence, in this case
$v=v_+$ we expect the formation of the upstream dispersive shock.
One should note that in the NLS dispersive hydrodynamics a general discontinuity
resolves into a certain combination
of two waves: dispersive shock(s) and/or rarefaction wave(s). The actual
combination depends on the relation between the specific initial jump values
for the density and velocity \cite{gk87,eggk95}. The closure conditions
enabling one to single out the unique combination
in our problem will be formulated in the next section.

Thus, one can suggest that in the two limiting cases $v=v_{\pm}$ the full solution can be
built of two parts: the steady hydraulic transition solution defined within the spatial
range of the potential barrier
and the unsteady dispersive shock/rarefaction wave combination at one of the sides
of the barrier.
Since  the dispersive shock is generated essentially  outside the spatial range of
the potential, one can use the potential-free NLS equation
for its description and take advantage of the  theory  developed in
\cite{gk87,eggk95} on the basis of the original Gurevich-Pitaevskii
approach using the Whitham method of slow modulations \cite{gp74,whitham74}.
This modulation theory  of the NLS dispersive shocks was recently applied
to the description of dispersive shocks in freely expanding BECs in \cite{kgk04,hoefer}.

\begin{figure}[bt]
\begin{center}
{\includegraphics[width= 7cm]{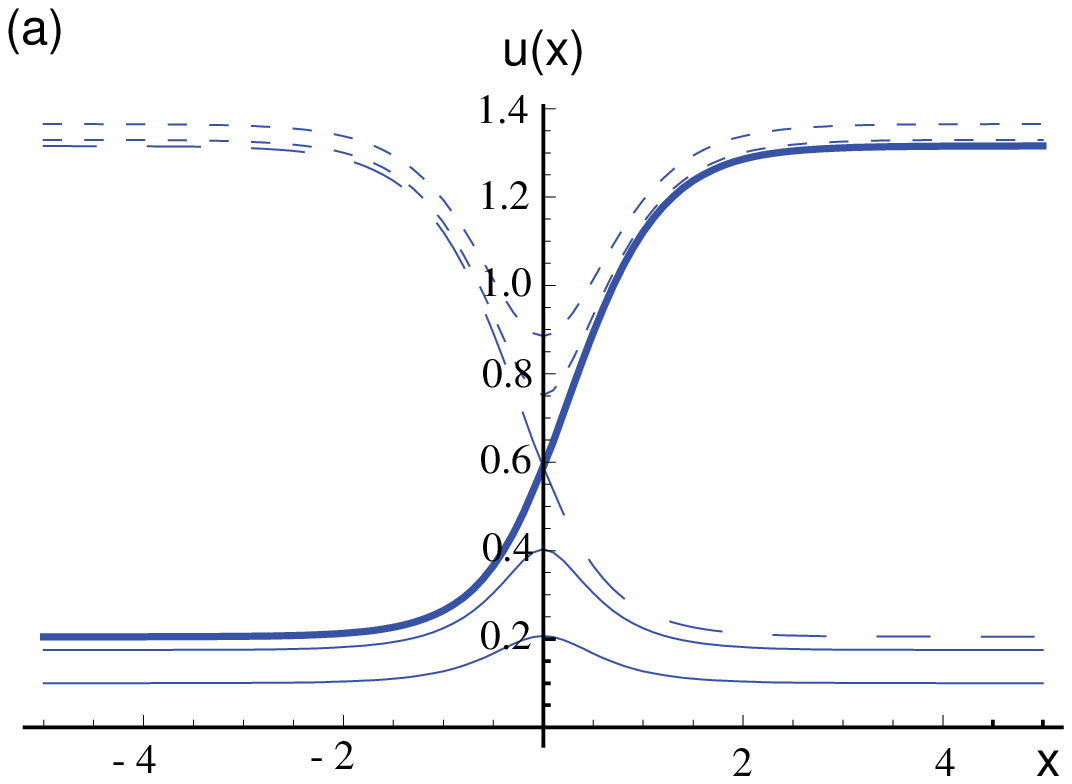}} \qquad
{\includegraphics[width= 7cm]{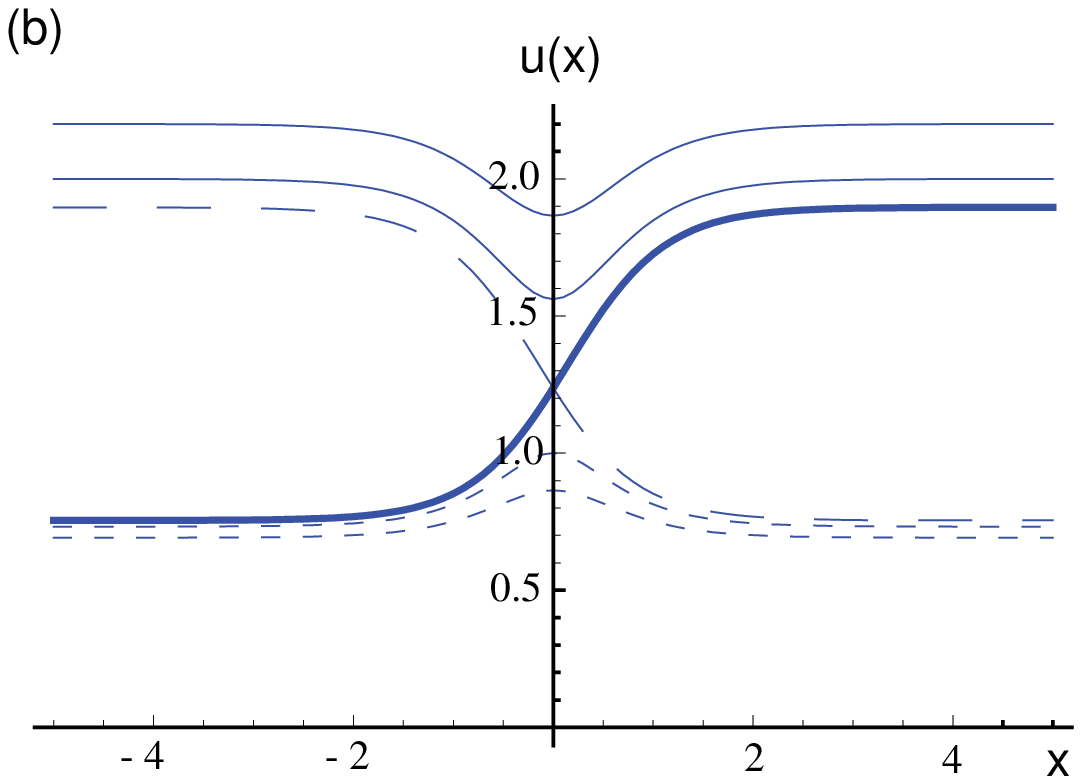}}
\caption{a) subcritical hydraulic solutions for $u$ at different
values of $v<v_-$ (thin solid lines) and  a separatrix solution at $v=v_-$ (thick solid line);
b) supercritical hydraulic solution at different values of $v>v_-$ (thin solid lines)
and separatrix solution at $v=v_+$ (thick solid line).  Short-dashed lines: non-physical branches. Long-dashed line: non-physical separatrix.}
\end{center}\label{fig4}
\end{figure}

Generally we are interested in the flow corresponding to the {\it transcritical} region
\begin{equation}\label{6-2}
    v_-\leq v\leq v_+
\end{equation}
so one can expect formation of both upstream and downstream discontinuities with their
further dispersive resolution into the
undisturbed states at $x\to\pm \infty$. To be precise, we expect a subcritical jump upstream,
a supercritical jump downstream and the
exact criticality at the top of the potential.
One should emphasize that the suggested description of the transcritical BEC
flow as a combination
of local hydraulic solution connected with the  equilibrium state at infinities
by unsteady dispersive shock(s)
is an {\it assumption} which should be  validated by the comparison of the
analytical solution with direct numerical simulations for a range of the oncoming
flow/potential barrier parameters. As we shall see, this assumption works very
well when the oncoming flow velocity is not too close to the transcritical region
boundaries $v_{\pm}(V_m)$. The reason for that is clear: if the speed
$v$ is close to one of its critical values $v_{\pm}(V_m)$, the  characteristic
relaxation time is expected to be very large (see \cite{brachet} for the
bifurcation scaling analysis in the case of the delta-functional potential)
so the local transcritical steady flow does not establish in finite time.
Nevertheless, it is instructive to perform the analysis, based on the above
assumption, for the whole range of velocities in the interval $[v_-, v_+]$
and then see how well this approximation works for different parameter values.

We denote the values of the density and
velocity at the boundaries $x=\pm l$ of the hydraulic solution  as
\begin{equation}\label{7-1}
\rho(-l)  \equiv  \rho^u  ,\quad u(-l) \equiv  u^u \quad\text{(upstream the barrier)}
\end{equation}
and
\begin{equation}\label{7-2}
     \rho(l)  \equiv  \rho^d  ,\quad u(l) \equiv u^d\quad\text{(downstream the barrier)}\, .
\end{equation}
We assume that the potential maximum is located at $x=0$, i.e. $V_x(0)=0$.
The transcritical hydraulic solution of our interest is then distinguished by
two sets of conditions:
\begin{equation}\label{crit}
\hbox{(i)  } \quad   u_x \ne 0 \ \ \hbox{at} \ \ x=0  ;
\end{equation}
\begin{equation}\label{ineq1}
\hbox{(ii)}  \quad u^u \le u_m, \quad \rho^u \ge \rho_m \  \hbox{ and} \quad
u^d \ge u_m, \quad \rho^d \le \rho_m  ,
\end{equation}
where $u_m, \rho_m$ are the values of $u$ and $\rho$  at $x=0$.  As we shall see the
condition (i) actually coincides with the requirement that
$u_m$ is equal to the local sound velocity at the top of the potential (exact criticality).

Now we take the solution of the
hydraulic equations (\ref{4-2}) in the general form
\begin{equation}\label{7-3}
    \rho u=c_1,\quad \tfrac12u^2+\rho+V(x)=c_2,
\end{equation}
$c_{1,2}$ being arbitrary constants, so that instead of (\ref{4-4}) we obtain
\begin{equation}\label{7-4}
    c_2-\frac{u^2}2-\frac{c_1}u = V(x) \, .
\end{equation}
Differentiating (\ref{7-4}) we obtain
\begin{equation}\label{}
\left (u -\frac{c_1}{u^2} \right ) u_x=V_x \, .
\end{equation}
Applying condition (\ref{crit})  we get
\begin{equation}\label{7-5}
    c_1=u_m^3 \, .
\end{equation}
Combining this relation with the first integral (\ref{7-3}) applied to
the point $x=0$ yields
\begin{equation}\label{7-6}
    \rho_m=u_m^2 \, ,
\end{equation}
which simply means that the local sound velocity at $x=0$ is
$\sqrt{\rho_m}=u_m$.

Substituting (\ref{7-5}), (\ref{7-6})  into  the second integral (\ref{7-3})
gives the value of $c_2$:
\begin{equation}\label{7-7}
    c_2=\frac32u_m^2+V_m.
\end{equation}
On the other hand, the integrals (\ref{7-3}) applied to the upstream
and downstream boundaries $x= \pm l$ where $V(x)\to0$ yield the relations
\begin{equation}\label{8-1}
    \rho^{u,d}u^{u,d}=u_m^3,\quad \tfrac12(u^{u,d})^2+\rho^{u,d}=\tfrac32u_m^2+V_m.
\end{equation}
Eliminating $u_m$  we get three equations
\begin{equation}\label{8-2}
   \rho^{u}u^{u} =  \rho^{d}u^{d}\, , \quad  \tfrac12(u^{u})^2+\rho^{u} =
   \tfrac12(u^{d})^2+\rho^{d}\, ,
\quad     \tfrac12(u^{u})^2+\rho^{u}-\tfrac32(\rho^{u}u^{u})^{2/3}=V_m
\end{equation}
for four quantities $\rho^{u,d},\,u^{u,d}$. Obviously, the last equation
can be replaced by
\begin{equation}\label{8-2a}
    \tfrac12(u^{d})^2+\rho^{d}-\tfrac32(\rho^{d}u^{d})^{2/3}=V_m,
\end{equation}
that is $u^u,\,\rho^u$ and $u^d,\,\rho^d$ satisfy the same set of equations.
One more equation is needed for closing the system, and this
will be obtained in the next subsection.

Once $u^{u,d}$ and $\rho^{u,d}$ are found,  the integration
constants $c_1$ and $c_2$ in (\ref{7-3})  are expressed as
\begin{equation}\label{10-1}
    c_1=\rho^{u,d}u^{u,d}, \ \ c_2=
    \tfrac12(u^{u,d})^2+\rho^{u,d}.
\end{equation}
Hence, the  transcritical hydraulic
solution (\ref{7-3}) is given by
\begin{equation}\label{10-2}
\tfrac12u^2+\frac{\rho^{u,d}u^{u,d}}u+V(x)=\tfrac12(u^{u,d})^2+\rho^{u,d},
\end{equation}
\begin{equation}\label{10-3}
    \tfrac12\left(\frac{\rho^{u,d}u^{u,d}}{\rho}\right)^2+\rho+V(x)=
    \tfrac12(u^{u,d})^2+\rho^{u,d}.
\end{equation}
At the boundaries $v=v_{\pm}$ of the transcritical region, where either
$\rho^u=1,\,u^u=v_-$ or $\rho^d=1,\,u^d=v_+$, equations (\ref{10-2}), (\ref{10-3}),
reduce to  (\ref{5-7}), (\ref{5-8}) respectively
as it should be.

\subsubsection{Closure conditions}
To get the closure condition for (\ref{8-2}) we shall take advantage of
the transition condition across the dispersive shock
which is most conveniently formulated in terms of the Riemann invariants of the
dispersionless hydrodynamic system
associated with the NLS equation. This system is obtained from (\ref{3-3})
by removing the dispersion and potential terms,
\begin{equation}\label{8-3}
    \begin{split}
    \rho_t+(\rho u)_{x}=0,\\
    u_t+uu_{x}+\rho_{x}=0 \, ,
    \end{split}
\end{equation}
and is equivalent to the classical shallow-water equations.
Introducing the Riemann invariants
\begin{equation}\label{8-5}
    \la_{\pm}=\tfrac12u\pm\sqrt{\rho},
\end{equation}
we represent (\ref{8-3}) in the diagonal form
\begin{equation}\label{8-4}
    \frac{\prt\la_{+}}{\prt t}+\tfrac14(3\la_++\la_-)\frac{\prt\la_{+}}{\prt x}=0,\quad
    \frac{\prt\la_{-}}{\prt t}+\tfrac14(\la_++3\la_-)\frac{\prt\la_{-}}{\prt x}=0 \, .
\end{equation}
In the problem of the decay of an initial discontinuity in the NLS
hydrodynamics the transition across the dispersive shock is characterized by the
zero jump of one of the hydrodynamic Riemann invariants  (\ref{8-5}) \cite{gk87,eggk95}.
It should be stressed that this condition is not a small-amplitude
approximation of the classical shock jump condition but rather is a non-perturbative
consequence of the data transfer along characteristics of the associated
Whitham modulation system describing dispersive shock region \cite{el05}.
It will transpire later that
in the present problem of the right-propagating BEC through a potential
barrier at rest, the ``conserving'' invariant is $\lambda_+$ (this corresponds to the waves
propagating {\it to the left} which might seem  somewhat counter-intuitive
with regard to the downstream dispersive shock).
\begin{figure}[bt]
\begin{center}
\includegraphics[width=10cm,height=6cm,clip]{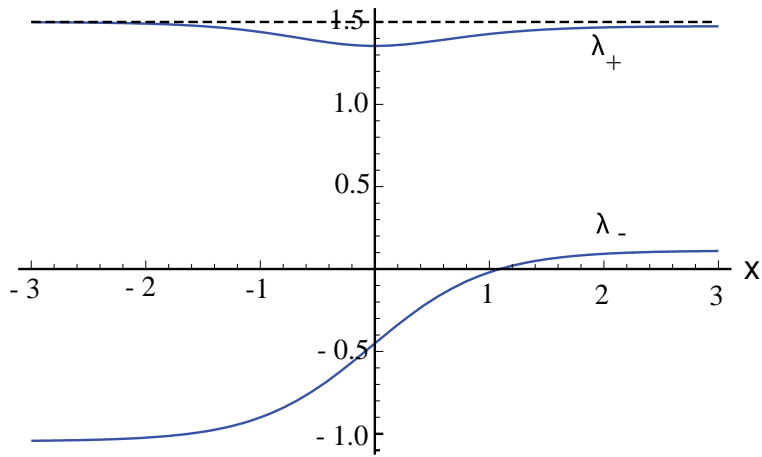}
\caption{Riemann invariants $\la_{\pm}$ as functions of $x$ in the
hydraulic solution (\ref{10-2}), (\ref{10-3}) for the potential  (\ref{6-1}) with $V_m=0.5$.
Note: $\la_+^u \approx \la_+^d$.}
\end{center}\label{fig5}
\end{figure}

The upstream dispersive shock is adjacent to the oncoming flow
$\rho=1$, $u=v$ so  the Riemann invariant condition yields
the relationship
\begin{equation}\label{8-6}
    \tfrac12u^u+\sqrt{\rho^u}=\tfrac12v+1 \, ,
\end{equation}
which closes the system (\ref{8-2}). We also recall that we are
interested in the solution satisfying condition (\ref{ineq1})
which allows one to identify the roots of (\ref{8-2}), (\ref{8-6})
as upstream and downstream ones. Now all four values
$\rho^{u,d}, u^{u,d}$ are fixed in terms of $V_m$, $v$, which,
in particular, implies that the downstream discontinuity
generally cannot be resolved by a single dispersive shock but
requires an additional right-propagating rarefaction wave to
adjust to the undisturbed flow at $+ \infty$.
While the introduction of this rarefaction wave does not
present one with a serious technical problem, we shall see
that in practical terms this additional complication
is not actually necessary.

First we note that,  since the matching of the (external)
hydrodynamic solution and the (internal) modulation solution
describing dispersive shock is most naturally  formulated in
terms of Riemann invariants (see \cite{gk87,eggk95}
for details) it is instructive to plot the transcritical
hydraulic solution
(\ref{7-3}),  (\ref{7-5}), (\ref{7-7}) in terms of $\lambda_{\pm}$
(the physical branch of the solution is selected using inequalities (\ref{ineq1})).
The plot in Fig.~5 suggests that for weak enough potentials one
can neglect the change of $\la_+$ across the barrier, which would
remove the necessity of introducing
an additional rarefaction wave. To justify this supposition we
first  eliminate $\rho^{u,d}$ from the second equation  (\ref{8-1}) to represent it in the form
\begin{equation}\label{1}
    \frac{(u^{u,d})^2}2+\frac{u_m^3}{u^{u,d}}-\frac32u^2_m=V_m.
\end{equation}
Introducing the normalized quantities $\nu_{\pm}=u^{d,u}/u_m$ we  represent  (\ref{1}) as
\begin{equation}\label{eq:1a}
\nu_{\pm}^2+\frac{2}{\nu_{\pm}}-3= \alpha,
\end{equation}
where $\alpha = 2V_m/u_m^2$.
Now let us suppose that $V_m\ll1$, which implies $\nu_{\pm}\approx1$,
$u \approx u_m\approx 1$, $\alpha \ll 1$.
Then from  (\ref{eq:1a}) we get the expansion
\begin{equation} \label{vapprox1}
\nu_{\pm}=1\pm \left( \frac{2\alpha}{3} \right)^{1/2} +\frac{2\alpha}{9}
+c_{\pm}\alpha^{3/2}+ \dots\, ,
\end{equation}
i.e. the controlling small parameter is again $\epsilon \sim V_m ^{1/2}$
(note that the coefficients $c_\pm$ in (\ref{vapprox1}) will
not contribute to the result so we do not present them explicitly).
We now consider the boundary values of the Riemann invariant $\lambda_+$ (\ref{8-5})
\begin{equation}\label{eq:3}
\lambda^{u,d}_+=\frac12u^{u,d}+\sqrt{\rho^{u,d}} = \frac12u^{u,d}
+\frac{u_m^{3/2}}{(u^{u,d})^{1/2}} \, .
\end{equation}
Again, normalizing, $\Lambda^{u,d}=\lambda_+^{u,d}/u_m$, and expanding for small $V_m$ we obtain
\begin{equation}\label{eq:4}
\Lambda^{d,u}= \frac{1}{2} \nu_{\pm}+\frac{1}{\nu_{\pm}^{1/2}} =
\frac{3}{2}+\frac{\alpha}{8} \mp \frac{1}{48 \sqrt{3}} \alpha^{3/2} \dots
\end{equation}
Now, taking into account that $u_m=1$ to leading order, we get
\begin{equation}\label{delta}
\delta = \lambda_+^{u} - \lambda_+^{d} =\left (\frac{V_m}{6}\right) ^{3/2}+\cdots
\end{equation}
Thus, for weak potentials the jump of the Riemann invariant $\la_+$ across
the potential has the third order in the controlling small parameter.
This result has certain analogy with the classical result from the shock
wave theory which reads that  the relevant
Riemann invariant has just the third order jump across the weak shock
\cite{CF48,LL6}. The coefficient $(1/6)^{3/2}$ before $V_m^{3/2}$ suggests  that
one can neglect the jump of $\la_+$ even for the  potentials of moderate
strength (of course, provided that the coefficients for the successive powers of
the small parameter $V_m^{1/2}$
in the expansion (\ref{delta}) are reasonably small). Indeed, for our potential
(\ref{6-1}) with $V_m=0.5$ we have $\la_+^{u,d} \approx 1.5$ while
according to (\ref{delta}) $\delta \approx 0.025$ i.e. just about $1.7\%$,
which is confirmed by the numerical transcritical hydraulic solution shown in Fig.~5.
So, for weak to moderate potentials  one can safely assume that the value of
the Riemann invariant $\lambda_+$ is preserved across the potential, which
allows one to use an additional closure condition
\begin{equation}\label{9-1}
    \tfrac12u^u+\sqrt{\rho^u}= \tfrac12u^d+\sqrt{\rho^d}=\tfrac12 v +1 \, .
\end{equation}
Relation (\ref{9-1})  is asymptotically consistent with exact conditions
(\ref{8-2}), (\ref{8-6}) and  will be especially useful in our further
consideration of the dispersive resolution
of the downstream discontinuity.  We also note that an immediate
implication of (\ref{9-1}) is that both upstream and downstream
dispersive shocks  are ``based''  on the same family of characteristics
corresponding to left-propagating hydrodynamic simple waves, which will
be essential for the modulation solution in the subsequent sections.

\subsubsection{Weak potentials: explicit formulae}

Using asymptotic closure conditions (\ref{9-1}) one can obtain simple approximate
explicit expressions  for $\rho^{u,d}$, $u^{u,d}$ in terms
of $v, V_m$, which will be useful later.  We use (\ref{9-1}) to eliminate
$\rho^{u,d}$ from  (\ref{8-2}) to obtain a single
equation for $w=u^{u,d}$,
\begin{equation}\label{9-2}
    \frac{w^2}2+\left(\frac{v-w}2+1\right)^2-\frac32\left[w
    \left(\frac{v-w}2+1\right)^2\right]^{2/3}=V_m.
\end{equation}
This equation has two roots,  the larger one corresponds to $u^d$
and the smaller one to $u^u$ (see (\ref{ineq1})).

Equation (\ref{9-2}) can be solved approximately for $V_m\ll1$.
It is easy to see that for $V_m=0$ this equation is satisfied if $w=1+(v-w)/2$ i.e.
$w=(v+2)/3$. In the next approximation we obtain
\begin{equation}\label{9-4}
    u^u=1+\frac13(v-1)-\sqrt{\frac{2V_m}3},\quad u^d=1+\frac13(v-1)+\sqrt{\frac{2V_m}3},
\end{equation}
where we have used that $v$  takes its values in the transcritical region  (see (\ref{5-6}))
\begin{equation}\label{9-6}
    1-\sqrt{\frac{3V_m}2} <  v < 1+\sqrt{\frac{3V_m}2}.
\end{equation}
Respectively, with the same accuracy we get from (\ref{9-1})
\begin{equation}\label{9-5}
    \rho^u= 1+ \frac{2}{3}(v-1) + \sqrt{\frac{2V_m}3},\quad
    \rho^d= 1+ \frac{2}{3}(v-1) - \sqrt{\frac{2V_m}3},
\end{equation}
Using (\ref{9-4}), (\ref{9-5}) we calculate the upstream and downstream values
of the Riemann invariant $\la_-= \tfrac 12 u - \sqrt{\rho}$,
which undergoes discontinuities at both edges of the transcritical hydraulic solution (see Fig.~5),
\begin{equation}\label{la-}
\la_-^u=\la_-(-l)= - \frac{1}{2} - \frac{1}{12}(v-1) - \sqrt{\frac{2V_m}3} \, ,
\qquad \la_-^d= \la_-(l)=- \frac{1}{2} - \frac{1}{12}(v-1) + \sqrt{\frac{2V_m}3}
\end{equation}
Obviously, $\la_-^u < \la_-^{\infty}$,  $\la_-^d > \la_-^{\infty}$, where
$\la_-^{\infty}=v/2-1$ is the value of $\la_-$ for the undisturbed flow at infinity.

\subsection{Resolution of downstream and upstream discontinuities}

Summarizing the results of the previous section, we have the following  values for
the Riemann invariants $\la_{\pm}$ at
the boundaries $x=\pm l$ of the hydraulic transition and at $x=\pm \infty$:
\begin{equation}\label{la+}
\la_+(-\infty)=\la_+^d=\la^u_+=\la_+(+\infty)=\tfrac 12 v +1 \, ,
\end{equation}
\begin{equation}\label{la-1}
\la_-(-\infty) = \la_-(+\infty)=\tfrac12 v-1\, ,
\end{equation}
\begin{equation}\label{la-2}
\la_-^d=\tfrac{1}{2}u^d -\sqrt{\rho^d} \, , \qquad   \la_-^u=\tfrac{1}{2}u^u -\sqrt{\rho^u}
\end{equation}
From now on we consider the values $\la_-^{u,d}(v, V_m)$ as known.
Thus, there are upstream and downstream discontinuities in $\la_-$
and no discontinuities in $\la_+$.

As was mentioned above, the upstream and downstream discontinuities are resolved
through the generation of  dispersive shock waves which can be described using
modulated periodic solutions of the defocusing NLS equation (\ref{3-1})
without the barrier potential $V(x+vt)$. The potential term can be neglected  because the
shock resolution occurs essentially outside the potential range $-l <x<l$.
The potential-free NLS equation is Galilean invariant and hence it preserves
its form (up to inessential in our case phase factor in $\psi$-function)
after the transformation to the reference frame with the barrier at rest which
is used in our calculations.

The periodic solution of the NLS equation can be written in terms of
the condensate density as (see, e.g., \cite{gk87,kamch2000})
\begin{equation}\label{11-1}
    \rho(x,t)=\tfrac14(\la_4-\la_3-\la_2+\la_1)^2+(\la_4-\la_3)(\la_2-\la_1)
    \mathrm{sn}^2(\sqrt{(\la_4-\la_2)(\la_3-\la_1)}\,\xi,m).
\end{equation}
Here
\begin{equation}\label{U}
   \xi=x-Ut,\quad U=\tfrac12\sum_{i=1}^4\la_i,\quad
    m=\frac{(\la_2-\la_1)(\la_4-\la_3)}{(\la_4-\la_2)(\la_3-\la_1)},
\end{equation}
$U$ being the phase velocity and $m$ the elliptic modulus, $0 \le m \le 1$.
The condensate velocity $u$ is expressed in terms of density as
\begin{equation}\label{11-2}
    u(x,t)=U - \frac{C}{\rho(x,t)},
\end{equation}
where
\begin{equation}\label{11-4}
    C=\tfrac18(-\la_1-\la_2+\la_3+\la_4)(-\la_1+\la_2-\la_3+\la_4)
    (\la_1-\la_2-\la_3+\la_4)\, .
\end{equation}
The wavelength is expressed in terms of $\la_j$'s as
\begin{equation}\label{L}
L=\frac{2{K}(m)}{\sqrt{(\la_4-\la_2)(\la_3-\la_1)}},
\end{equation}
$K(m)$ being the complete elliptic integral of the first kind.

The parameters $\la_1\leq\la_2\leq\la_2\leq\la_4$ change slowly through the
dispersive shock and their evolution is governed by the Whitham
modulation equations
\begin{equation}\label{11-5}
    \frac{\prt\la_i}{\prt t}+v_i(\la)\frac{\prt\la_i}{\prt x}=0,\quad i=1,2,3,4,
\end{equation}
so $\la_j$'s are the Riemann invariants of the Whitham
equations. Here $v_i(\la_1,\ldots,\la_4)$, $i=1,2,3,4,$ are the characteristic
velocities which can be expressed in terms of the complete elliptic integrals
of the first and the second kind \cite{fl86,pavlov87}. One can use the
following convenient formula for the computation of $v_i$'s  \cite{gke92,kamch2000}
\begin{equation}\label{vi}
v_i= U-\frac 12 \frac{L}{\partial L/\partial \la_i} ,\quad
i=1,2,3,4.
\end{equation}

The waveform (\ref{11-1}) within the dispersive shock wave region gradually
changes from the vanishing
amplitude harmonic wave at one of the edges
to a dark soliton at the opposite edge so that generally the modulus $m$ runs
over the whole range from $0$ to $1$.
The Riemann invariants $\la_{j}$ of the Whitham equations  are matched with the
``external'' hydrodynamic
invariants $\la_{\pm}$ at some free boundaries  $x^{\pm}(t)$  defined by
the conditions $m=0$ (harmonic edge) or $m=1$ (soliton edge).
The specific matching conditions and relative position of the harmonic
and soliton edges depend on whether one considers the left- or right-propagating
dispersive shock (see \cite{eggk95}).
For the left-propagating case of our interest the matching conditions are
\begin{equation}
\begin{array}{l}
\la_2=\la_1\, , \  \ \la_4 = \la_+ , \  \   \la_3 = \la_- \quad \hbox{at} \ \ x=x^-(t);\\
\la_2=\la_3\, , \ \  \la_4 = \la_+ ,  \  \  \la_1 = \la_- \quad \hbox{at} \ \ x=x^+(t) .
\end{array}
\label{bc}
\end{equation}
\begin{figure}[bt]
\begin{center}
\includegraphics[width=10cm,clip]{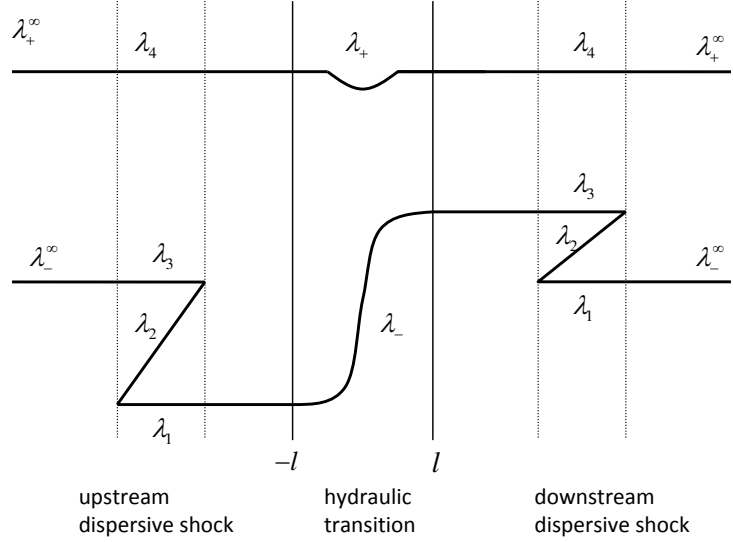}
\caption{Qualitative behavior  of the Riemann invariants $\la_j $ as functions
of $x$ in the full hydraulic/modulation solution.}
\end{center}\label{fig6}
\end{figure}
Since dispersive shocks expand with time,  their widths  at sufficiently
large $t$  are much greater than the range $l$ of the barrier
potential, so for $t \gg l$ one can assume the hydraulic solution to be
asymptotically localized at $x=0$ and use the similarity solutions of
the Whitham equations (\ref{11-5}) with only one variable changing
through each shock \cite{gk87} (note that  the formal condition of
applicability of the Whitham method itself to the decay of a step problem
is $t \gg 1$ --- see \cite{gp74,el05}).
From the matching conditions (\ref{bc}) it is clear that the varying invariant
should be $\la_2$. Thus, in each dispersive shock the invariants
$\la_1,\,\la_3,\,\la_4$ are constant and the invariant $\la_2$ varies
according to the similarity solution
\begin{equation}\label{12-2}
    v_2(\la_1,\la_2,\la_3,\la_4)=\frac{x}t,
\end{equation}
where
\begin{equation}\label{12-3}
    v_2(\la_1,\la_2,\la_3,\la_4)=\tfrac12\sum\la_i+\frac{(\la_3-\la_2)(\la_2-\la_1)K(m)}
    {(\la_3-\la_2)K(m)-(\la_3-\la_1)E(m)} ,
\end{equation}
$E(m)$ being the complete elliptic integral of the second kind.
The values of $\la_1,\la_3,\la_4$ in (\ref{12-3}) are fixed and
the Riemann invariant $\la_4$ must have the same constant value
\begin{equation}\label{}
\la_4=\tfrac12 v+1
\end{equation}
in both  downstream and upstream dispersive shocks since the value of
$\la_+$ is transferred through the hydraulic transition (see (\ref{la+})).
At the same time, the values of $\la_1$ and $\la_3$
are different upstream and downstream the barrier so we shall consider
the downstream and upstream dispersive shocks separately.
Before we proceed with the detailed analysis of the dispersive shocks
it is instructive to get a picture of qualitative behavior of the Riemann invariants in the
entire flow. For that, we use the matching conditions (\ref{bc}) and the
inequality $\partial \la_2 /\partial x >0$ following from the similarity modulation solution
(\ref{12-2}) and the general property $\partial v_i/\partial \la_i>0$
of the characteristic velocities (\ref{vi}).
As a result, one arrives at the scheme of the behavior of the Riemann
invariants sketched in Fig.~6.

\subsubsection{Downstream dispersive shock/soliton train}

In the downstream dispersive shock we have (see the matching conditions (\ref{bc}) and Fig.~6)
\begin{equation}\label{13-1}
    \la_1=\la_-^{\infty}=\frac{v}2-1,\quad \la_3=\la_-^d,\quad \la_4=\la_+^{\infty}=\frac{v}2+1,
\end{equation}
and $\la_2$ as a function of the self-similar variable $x/t$ is determined by
the equation
\begin{equation}\label{13-2}
    v_2(v/2-1,\la_2,\la_-^d,v/2+1)=\frac{x}t.
\end{equation}
In the linear limit $\la_2\to\la_1$ (i.e. $m \to 0$) we have
\begin{equation}\label{13-3}
    v_2(\la_1,\la_1,\la_3,\la_4)=\la_1+\tfrac12(\la_3+\la_4)+\frac{2(\la_3-\la_1)(\la_4-\la_1)}
    {2\la_1-\la_3-\la_4}\, .
\end{equation}
This velocity determines the speed $s^d_-$ of the trailing edge of the
downstream dispersive shock and it is not difficult to check using
formulae (\ref{9-4}), (\ref{la-})  that for $V_m\ll1$ it  is negative,
that is at least for weak potentials the trailing edge cannot be located
in the downstream region. Hence, in the
equation (\ref{13-2}) the variable $\la_2$ is limited from below by some cut-off value
$\la_2^*$,  and, as a result, the downstream shock gets attached to the
barrier at its boundary $x\approx l$ .
Then one can find $\la_2^*$ in the approximation described above
by taking $l/t\to0$ in Eq.~(\ref{13-2}), which yields
\begin{equation}\label{13-4}
    v_2(v/2-1,\la_2^*,\la_-^d,v/2+1)=0.
\end{equation}
Since $v_2$ has the meaning of nonlinear group velocity, equation (\ref{13-4})
can be interpreted as an asymptotic condition that the modulated wave
``stops'' at $x = 0$ and this is why it can be directly connected
(on the level of the Riemann invariants, i.e. in the ``averaged'' sense)
to the stationary hydraulic transition, which on the modulation length scale
can be viewed as a discontinuity located at $x=0$. In more precise terms, relation
(\ref{13-4}) means that for the solution under study the characteristic of
the Whitham equations  $dx/dt=v_2$
coincides with the characteristic $dx/dt=0$ of the dispersionless
equations (\ref{8-3})  in the hydraulic approximation at $x=0$ so the
modulation solution can be  ``terminated'' at $x=0$ and the free-boundary
matching conditions (\ref{bc}) at  the (nonexistent) trailing edge $x^-<0$
can be replaced with the
boundary conditions at $x=0$:
\begin{equation}\label{}
\lambda_3=\lambda^d_-, \ \ \la_4=\tfrac{1}{2}v+1 , \ \ \la_1=\tfrac{1}{2}v-1
\quad\hbox{at}\quad x=0 .
\end{equation}
The modulation solution  is then considered  in the upper right quarter
of $x,t$-plane, $x>0, t>0$.  One should mention that the similar situation
occurs in the problem of the transcritical shallow water fluid flow past
an obstacle described by the forced KdV equation
where the ``partial undular bore'' attached to the obstacle is generated
in the upstream flow \cite{gs86,s87}
(see also \cite{ms}).
Qualitative behavior of the Riemann invariants in the attached downstream
dispersive shock is shown in Fig.~7.
\begin{figure}
\begin{center}
\includegraphics[width=6cm,clip]{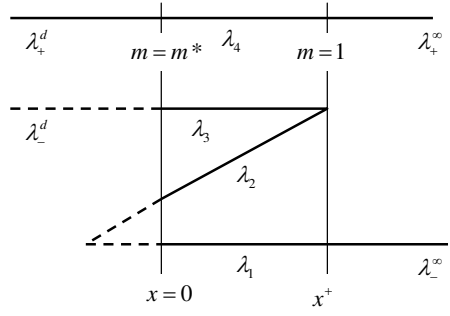}
\caption{ Qualitative  behavior  of the Riemann invariants in the partial
(attached) downstream dispersive shock.}
\end{center}\label{fig7}
\end{figure}
Equation (\ref{13-4}) determines $\la_2^*$, and therefore, via (\ref{U}),
the value of the modulus $m^*$ at which the downstream dispersive shock is generated
at $x=0$. This value is equal to
\begin{equation}\label{}
m^*=\frac{(\la_2^*-\tfrac 12 v +1)(\tfrac12v + 1 -\la_-^d)}
{(\tfrac12 v +1-\la_2^*)(\la_-^d-\tfrac 12 v +1)}.
\end{equation}
Within the partial dispersive shock the modulus changes in the interval $m^*  \le m \le 1$.
The dependence of the cut-off modulus $m^*$ on the BEC flow velocity $v$
is shown in Fig.~8a   for $V_m=0.5$.
One can see that for this case  the cut-off modulus ranges in the interval
$0<m^*\lesssim 0.75$ and for  the transcritical  flow velocities $v$ sufficiently close
to the lower  boundary $v_-$ one can treat the downstream dispersive shock
as a dark soliton train slowly propagating to the right relative to the barrier.
This is the dark soliton train that should remain within the finite
elongated BEC for some time after the potential has been swept and,
therefore, could be observed in the experiment. Therefore it is
instructive to study its parameters in more detail.

First we calculate the downstream soliton emission rate.
This can be done using the formula
\begin{equation}\label{f*}
f^*= U^* /L^* \, ,
\end{equation}
where
\begin{equation}\label{}
\begin{split}
U^*=\tfrac{1}{2}(\la_1+\la_2^*+\la_3+\la_4) = \tfrac{1}{2}(v+\la^d_-+\la_2^*)\, ,  \\
L^*=L(\la_1,\la_2^*,\la_3,\la_4)=\frac{2 K(m^*)}
{\sqrt{(\tfrac{1}{2}v+1-\la_2^*)(\la_d^--\tfrac{1}{2}v +1)}}
\end{split}
\end{equation}
are the phase velocity  (\ref{U}) and the wavelength (\ref{L}) respectively,
calculated at the soliton train generation point $x=0$.
\begin{figure}
\begin{center}
\includegraphics[width=6cm,clip]{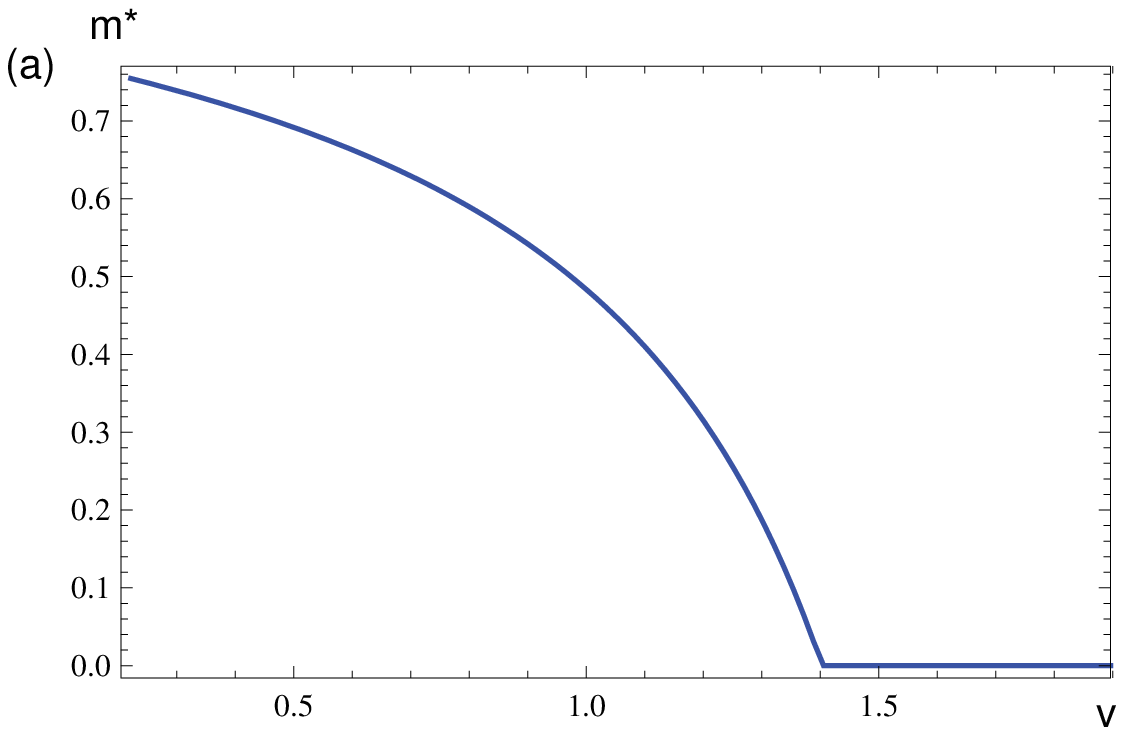}  \qquad  \qquad
\includegraphics[width=7cm,clip]{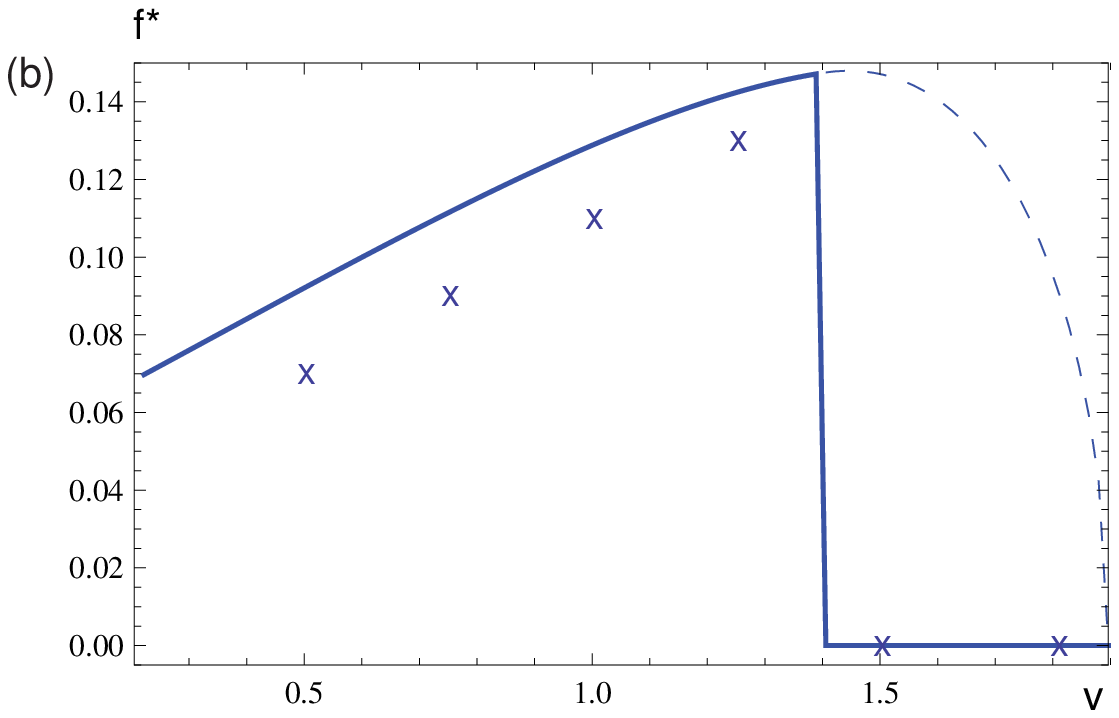}
\caption{a) Cut-off modulus $m^*$ in the downstream  dispersive shock (dark soliton train)  and
b)  Downstream dark soliton emission rate $f^*$ at the potential location vs oncoming BEC velocity $v$.
Both figures correspond to $V_m=0.5$.
At $v \approx 1.4$ the downstream dispersive shock wave detaches from the potential.
Crosses in (b) correspond to the numerical simulations data. The dashed line in (b)
corresponds to the frequency at the trailing edge of the detached downstream dispersive
shock. }
\end{center}\label{fig8}
\end{figure}

Dependence $f^*(v)$ for $V_m=0.5$ is shown in Fig.~8b where it is also compared
with the data from the direct numerical simulation using the potential (\ref{6-1})
with $V_m=0.5$, $\sigma=2$. Some features seen on this plot deserve  additional explanation.
Firstly we notice that the analytically obtained value for the soliton emission
frequency is finite for the flow velocities close to (but slightly greater than)
the lower transcritical boundary $v_- \approx 0.2$.
At the same time, the numerically obtained values of $f^*$ suggest that the rate
of soliton production  tends to zero as one approaches the lower transcritical
region boundary $v_- \approx 0.2$ (we did not observe any soliton generation at $v=0.25$
for the time range up to $t=50$).
The disagreement between the analytical solution and actual behaviour of the
soliton emission frequency is due to the failure, close to the transcritical region
boundary $v=v_-$, of our main assumption about the existence of the local
transcritical hydraulic solution forming discontinuities with the equilibrium
basic flow (see the discussion in Section B1). We note that the detailed analysis of the
dynamical scaling law for the soliton emission frequency for near-critical NLS flows
through short-range (delta-function) potentials was performed in \cite{brachet},
where the frequency was shown to vanish as $\delta ^{1/2}$, $\delta$
being the deviation of the controlling parameter (for instance, potential strength)
from its critical value.

The second ``non-standard'' feature in Fig.~8b is an abrupt change, back to zero,
of the emission frequency $f^*$ for
$v \approx 1.4$. This change does not constitute the cease of the soliton
generation downstream but simply reflects the fact that
the downstream dispersive shock gets detached from the obstacle potential for
velocities greater than some $v=v^*$, so there are no waves generated at the
potential location at $x=0$ (where the frequency $f^*$ is defined).
Indeed, as $v$ increases within the transcritical region $[v_-, v_+]$,
the speed $s^d_-$ of the trailing edge (computed formally by (\ref{13-3}))
can change the sign from minus to plus (see Fig.~9),  which implies that
for some $v=v^*$  the downstream dispersive shock must detach from the barrier.
The detachment threshold velocity is determined from the condition
\begin{equation}\label{}
s^d_-(v^*)=0 ,
\end{equation}
where $s_-^d(v)=v_2(m=0)$ is given by equation (\ref{13-3}).
For $V_m=0.5$ this velocity is
$v^* \approx 1.4$ (which can also be clearly seen in Figs.~8a,b).

One of the physical consequences of the downstream dispersive shock detachment
from the obstacle is the zero frequency of oscillations for the drag force
(see Section 2D below). At the same time, one should note that at the point of the
detachment, the amplitude of the dispersive shock vanishes, so the described
discontinuity in the emission frequency at the obstacle has no practical significance.
We stress that the solitons keep get generated downstream for
$v^*<v<v_+$, but the generation point---the trailing edge of the downstream
dispersive shock---now moves away from the potential with constant velocity
$s_-^d$. The corresponding wave frequency at the moving generation point is
shown in Fig.~8b by the dashed line.
\begin{figure}
\begin{center}
\includegraphics[width=6cm, clip]{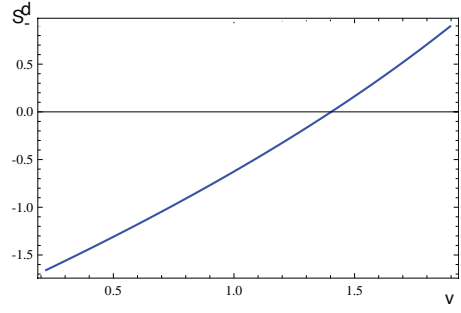}
\caption{Dependence of the downstream dispersive shock trailing edge
speed $s^d_-$ vs transcritical BEC velocity $v$ calculated by formula (\ref{13-3}).
The detachment point $v=v^* \approx 1.4$ is found from the condition  $s^d_-=0$.}
\end{center}\label{fig9}
\end{figure}
\begin{figure}
\begin{center}
\includegraphics[width=6cm,clip]{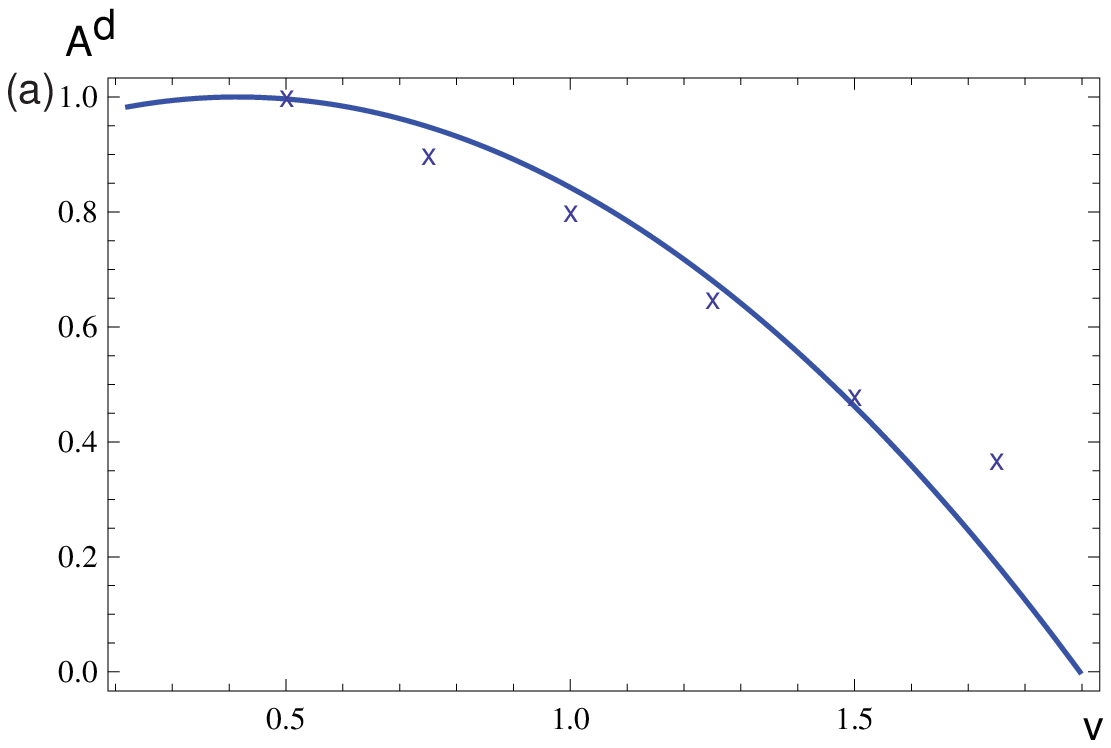}  \qquad  \qquad
\includegraphics[width=6cm,clip]{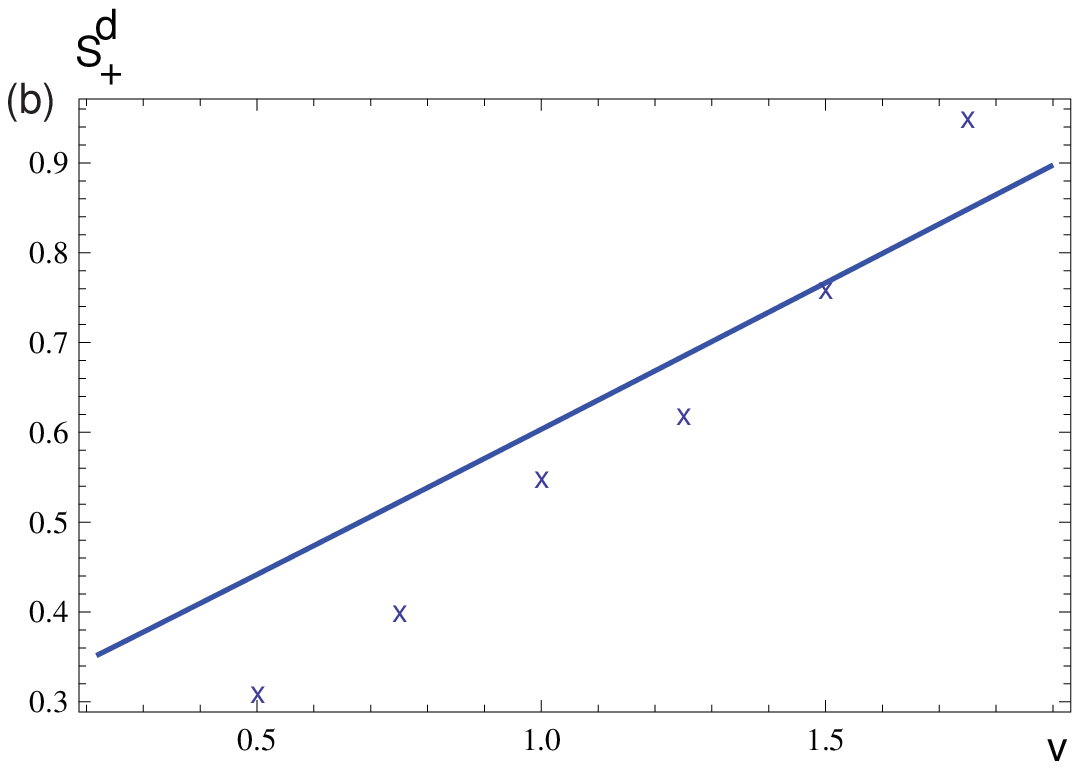}
\caption{Leading soliton parameters: a) amplitude $A^d$  and
b) speed $s^d_+ $ ---  in the downstream  dispersive shock
(dark soliton train)  vs oncoming BEC velocity $v$ for  $V_m=0.5$. Solid line: modulation solution; Crosses: direct numerical solution. }
\end{center}\label{fig10}
\end{figure}
\begin{figure}
\begin{center}
\includegraphics[width=6cm,clip]{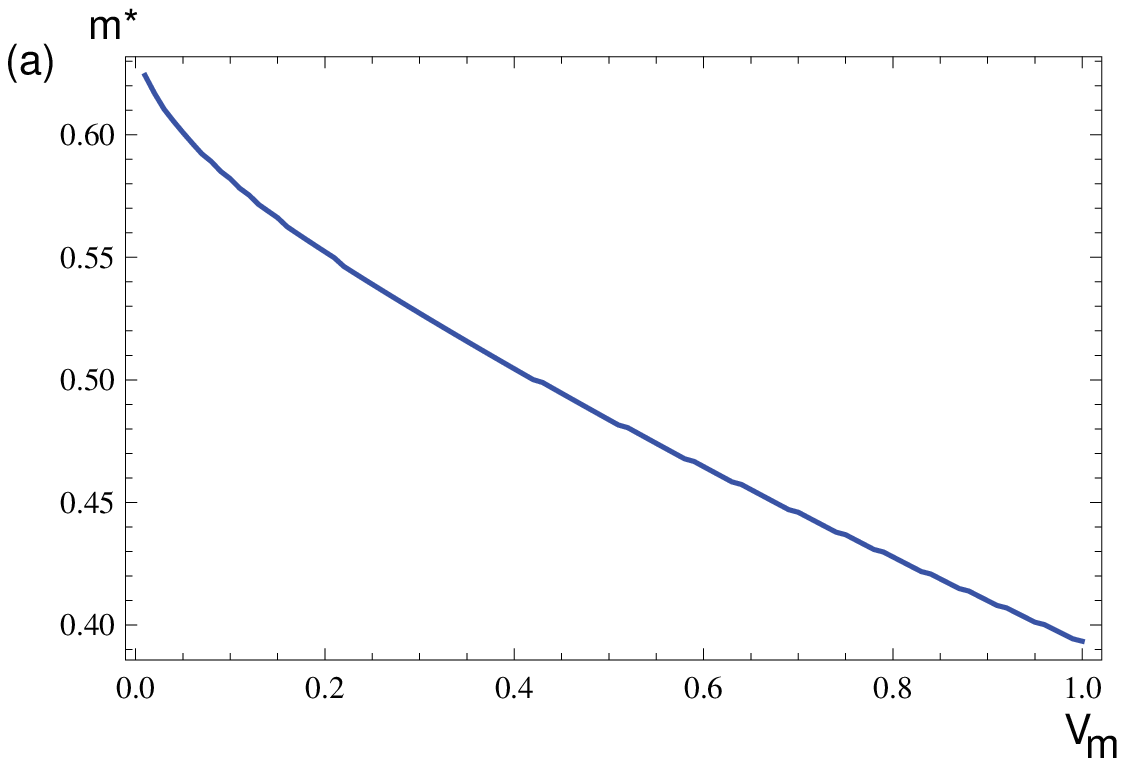}  \qquad  \qquad
\includegraphics[width=6cm,clip]{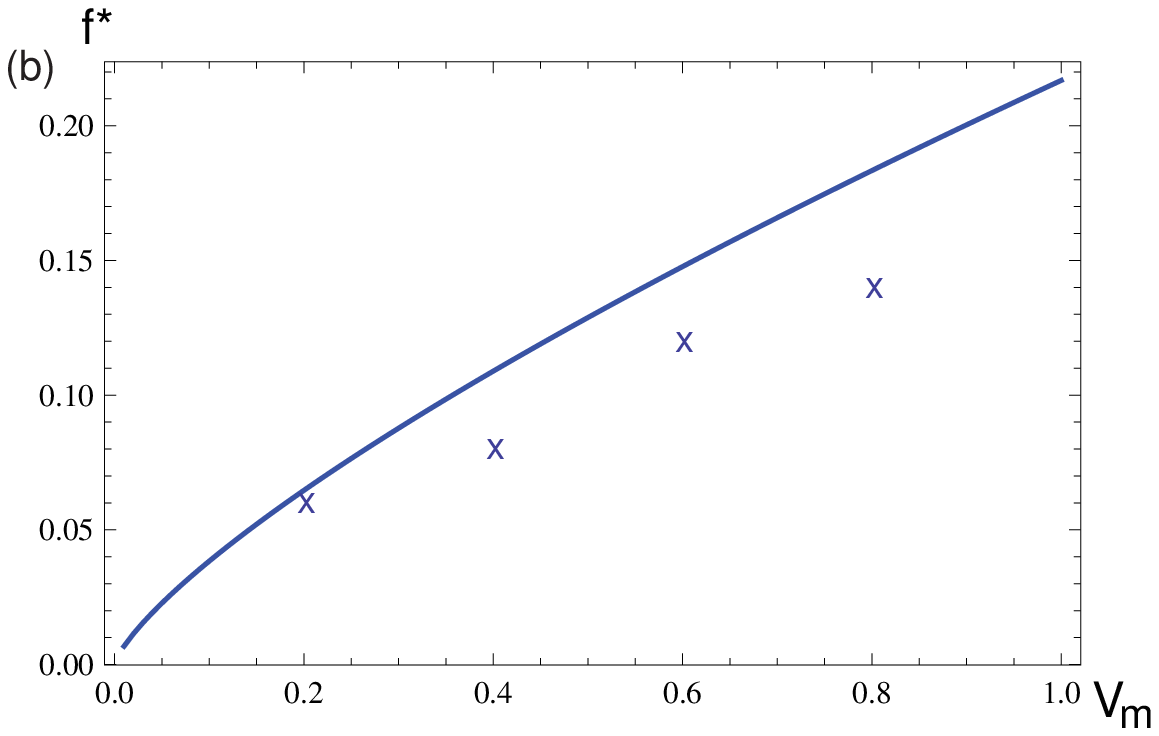}
\caption{a) Cut-off modulus $m^*$ in the downstream  dispersive shock  and
b) Downstream dark soliton emission rate $f^*$ vs potential strength  $V_m$.
Both figures correspond to $v=1.0$. Crosses in (b) correspond to the numerical simulations data. }
\end{center}\label{fig11}
\end{figure}
\begin{figure}
\begin{center}
\includegraphics[width=6cm,clip]{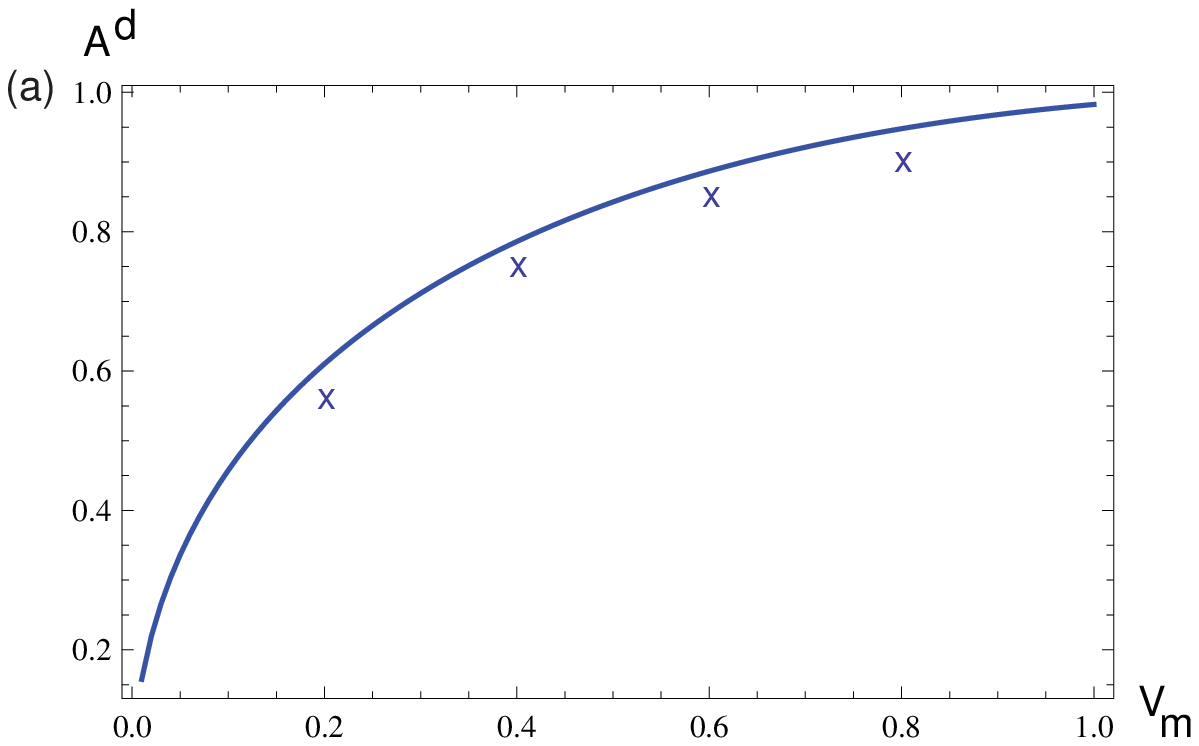}  \qquad  \qquad
\includegraphics[width=6cm,clip]{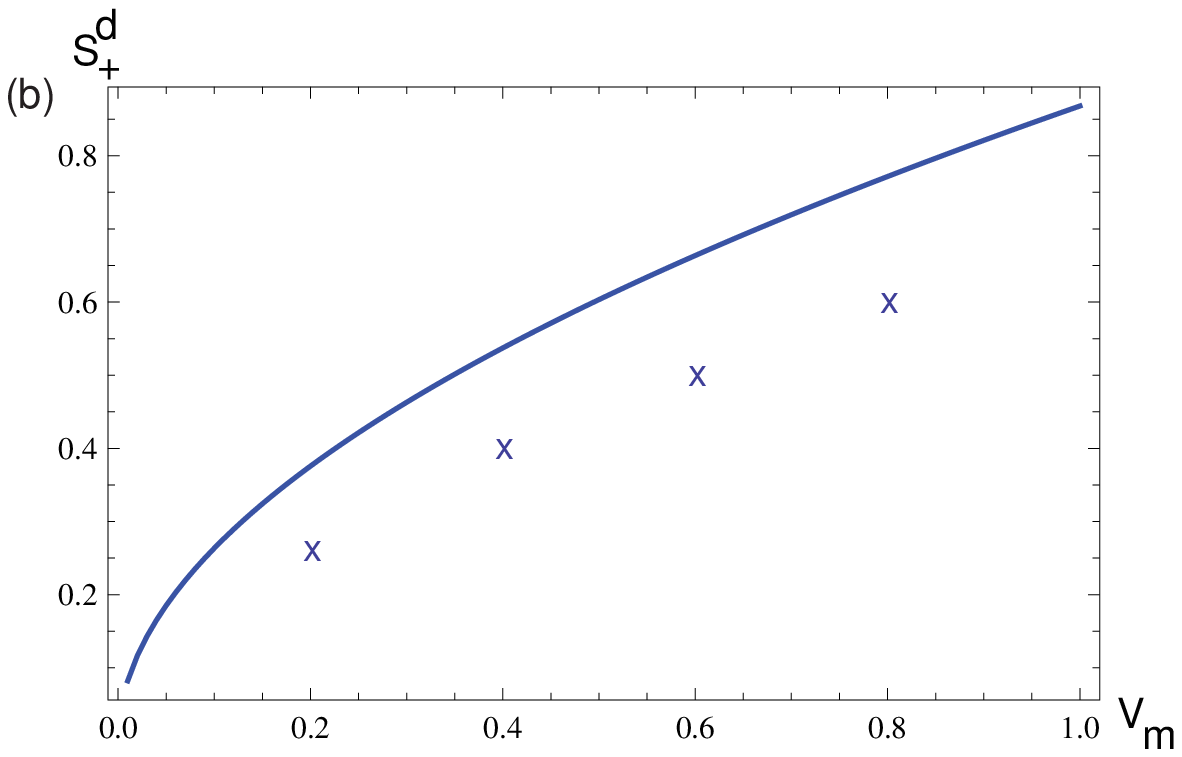}
\caption{Leading soliton parameters: a) amplitude $A^d$  and
b) speed $s^d_+ $  ---  in the downstream  dispersive shock
(dark soliton train)  vs potential strength $V_m$.
Both figures correspond to $v=1.0$. Solid line: modulation solution;
Crosses: direct numerical solution.}
\end{center}\label{fig12}
\end{figure}

The limit $\la_2\to\la_3$ ($m \to 1$) corresponds to the leading edge
$x^+$ of the downstream dispersive shock.
The amplitude of the leading soliton is given by
\begin{equation}\label{13-5}
    A^d=(\la_4-\la_3)(\la_3-\la_1)=(2+v-u^d)(u^d-v),
\end{equation}
and it moves with the velocity
\begin{equation}\label{13-6}
    s_+^d=v_2(\la_-^{\infty},\la_-^d,\la_-^d,\la_+^{\infty})=u^d-1.
\end{equation}
The dependencies $A^d(v)$ and $s_+^d(v)$ for $V_m=0.5$ are shown in Fig.~10
along with the corresponding numerical simulations data.
In Figs.~11, 12 we plot  the same set  of quantities:
$A^d, s_+^d, m^*$ and $f^*$---as functions of
the potential strength $V_m$  for fixed $v=1$. One can see that our solution
predicts the main physical parameters of the downstream wave quite well
provided the flow velocity is not too close to the critical values $v_{\pm}$
and the potential strength $V_m$ is not too large.

Summarizing, the downstream dispersive shock occupies the region
\begin{equation}\label{13-7}
    s_-^*<x<s_+^d\cdot t,
\end{equation}
where $s_-^*= 0$ for $v_-<v<v^*$ and  $s_-^*=s^d_-$ for $v^* <v < v_+$.

\subsubsection{Upstream dispersive shock}

The calculations in this case are very similar to those for the downstream dispersive shock.
Now the constant Riemann invariants are equal to
\begin{equation}\label{14-1}
    \la_1=\la_-^u,\quad \la_3=\la_-^{\infty}=v/2-1,\quad \la_4=\la_+^{\infty}=v/2+1,
\end{equation}
and $\la_2$ as a function of $x/t$ is determined by the equation
\begin{equation}\label{14-2}
    v_2(\la_-^u,\la_2,v/2-1,v/2+1)=x/t.
\end{equation}
The zero-amplitude trailing  edge with $\la_2\to\la_1$ propagates with the velocity
\begin{equation}\label{14-3}
    s_-^u=v_2(\la_-^u,\la_-^u,v/2-1,v/2+1)=2\la_-^u-\frac1{\la_-^u-v/2}
\end{equation}
which is always negative. The limit $\la_2\to\la_3$ corresponds to the leading (soliton) edge.
The amplitude of the leading soliton is equal to
\begin{equation}\label{15-1}
    A^u=2(v-u^u),
\end{equation}
and it moves with the velocity
\begin{equation}\label{15-2}
    s_+^u=v_2(\la_-^u,v/2-1,v/2-1,v/2+1)=\tfrac12(u^u+v-2).
\end{equation}
In the case of of small $V_m\ll1$ we get on using expansion (\ref{9-4}) that
\begin{equation}\label{15-3}
    s_+^u\cong\tfrac23(v-1)-\sqrt{\frac{V_m}6} \, ,
\end{equation}
which implies  that in the transcritical interval (\ref{9-6})
$s_+^u$ positive. Thus there exists the range of velocities $v$ for which the upstream
shock is also attached to the barrier and realized only partially.
In contrast to the attached downstream dispersive shock,
in the partial upstream wave the modulus $m$ varies between $0$ and
some cut-off value $m^{**}<1$, i.e. this wave can  be viewed as a
nonlinear oscillatory tail rather than solitary wave train.

The  value
$\la_2=\la_2^{**}$,  is found
from the equation
\begin{equation}\label{15-5}
    v_2(\la_-^u,\la_2^{**},v/2-1,v/2+1)=0,
\end{equation}
and determines the value of the cut-off modulus $m^{**}$ via (\ref{U}).
The threshold  velocity $v=v^{**}$ at which the upstream dispersive
shock gets attached to the potential is found from the
condition $s^u_+=0$ and is determined implicitly by the equation
\begin{equation}\label{15-6}
    u^u(v)=2-v,
\end{equation}
so that in the interval of velocities
\begin{equation}\label{15-7}
    v^{**}<v<v_+
\end{equation}
the upstream dispersive shock is attached to the barrier. Thus, the upstream shock
occupies the region
\begin{equation}\label{15-8}
    s_-^u\cdot t<x<s_+^{**}\cdot t,
\end{equation}
where $s_+^{**}= 0$ for the velocities $v$ in the interval (\ref{15-7})
and $s^{**}= s_+^u$ for $v_- < v < v^{**}$.

\subsection{Consolidated wave pattern}

Putting together the analytical results of this Section, we obtain full
asymptotic description of  the wave
pattern generated in the transcritical  1D  flow of a BEC past a wide penetrable
barrier. The pattern consists
of two dispersive shocks  propagating upstream and downstream the barrier and
connected with each other by the transcritical hydraulic transition
localized over the potential barrier spatial range. The behavior of the
density and velocity within the dispersive shocks  is obtained by the substitution of the
slowly varying similarity modulation solutions (\ref{13-1}), (\ref{13-2})
and (\ref{14-1}), (\ref{14-2}) for $\lambda_j$  into the
rapidly oscillating traveling wave solution (\ref{11-1}) and (\ref{11-2}).
One should note that the matching conditions (\ref{bc}) used in the construction
of the modulation solution only guarantee the continuous matching of the average
flow at the dispersive shock boundaries. To get exact matching
of the rapidly oscillating wave field within the dispersive shocks with the
constant (or slowly varying) flow outside, one should use the higher order
analysis.  The ``weak limit'' formulation of the dispersive shock problem
used above is  now well established owing to the studies based on the Lax-Levermore-Venakides
rigorous approach (see \cite{llv,jl99} and references therein).

The obtained combined modulated/hydraulic solution is shown in Fig.~13 at $t=30$
for the potential with $V_m=0.5$
and the oncoming flow velocity $v=1$ (phase adjustments within the accuracy
of the modulation theory are made  to
ensure continuity of the graph at the boundaries of the dispersive shocks).

We have also performed direct numerical integration the Gross-Pitaevskii equation (\ref{3-3}) with boundary conditions
(\ref{3-4}). The simulations were performed using two different methods: the classical finite difference explicit scheme \cite{ta84} and the quasi-spectral split step method \cite{jaru06}. Both methods gave the same results.
In Fig.~14 the numerical solution of the GP equation (\ref{3-3}) is plotted
for the potential (\ref{6-1}) with $V_m=0.5$ and $\sigma=2$.
One can see excellent agreement between the wave patterns in Figs.~14 and 15.
\begin{figure}[bt]
\begin{center}
\includegraphics[width=10cm, clip]{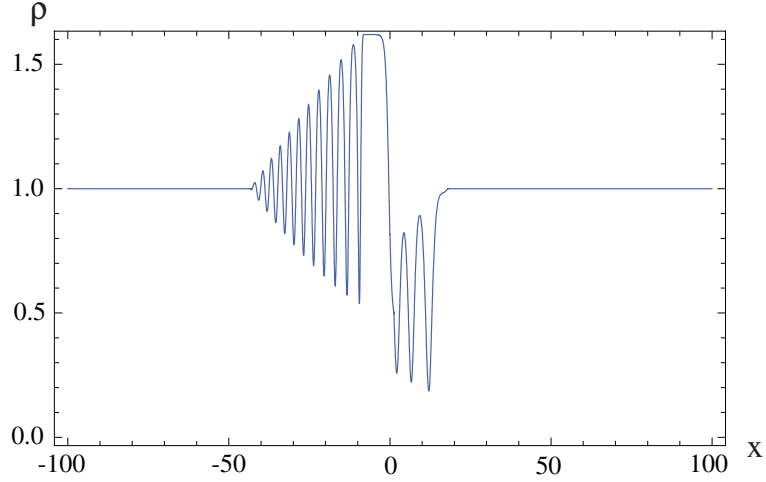}
\caption{Plot of the combined analytical (hydraulic + modulated oscillatory)
solution for the condensate density distribution in the wave pattern
generated by a BEC flow with $v=1$ through the potential barrier
with $V_m=0.5$  at $t=30$.}
\end{center}\label{fig13}
\end{figure}
\begin{figure}[bt]
\begin{center}
\includegraphics[width=11.5cm,clip]{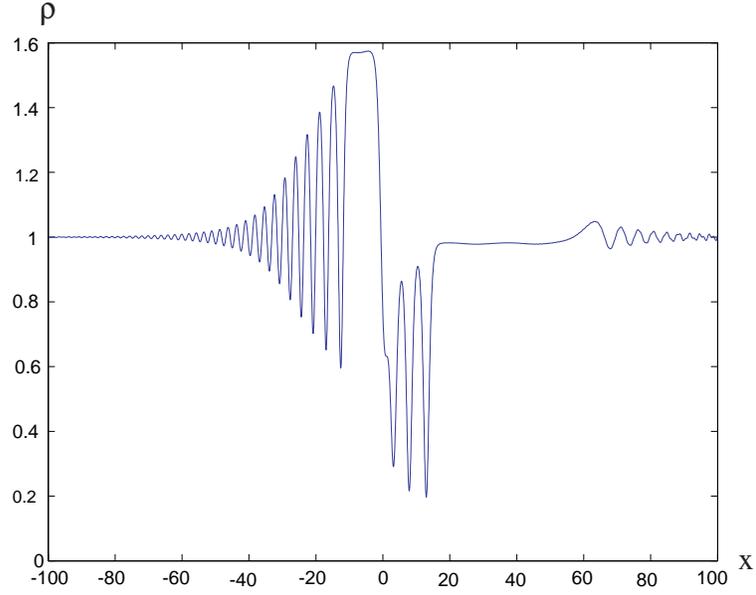}
\caption{Numerical simulation of the condensate density distribution
as a function of the space coordinate $x$ in the wave pattern
generated by a BEC flow with $v=1$ through the potential barrier (\ref{6-1}) with $V_m=0.5$, $\sigma=2$. The evolution time is equal to $t=30$.}
\end{center}\label{fig14}
\end{figure}

An additional small wave located about $x=60$ in the numerical solution (Fig.~14)
is due to the generation of a small-amplitude right-propagating wave packet
formed from the Bogoliubov linear waves created by a switching on the obstacle's
potential (see \cite{bks08}). The group velocity $v_g=d\omega/dk$ of the Bogoliubov
waves obeying the dispersion relation $\omega(k)=k\sqrt{1+k^2/4}$ is always greater than the
sound speed $c_s=\sqrt{\rho}=1$ and tends to this value in the long wavelength limit
$k\to0$. Taking into account that this wave packet is convected by the flow with
velocity $v$, we find that the wave packet occupies the region $x>(c_s+v)\cdot t$
and its predicted position is about $x>60$ for the chosen parameters with
$c_s=v=1$ and $t=30$ in agreement with the numerical results.
Besides this wave packet, one can notice a tiny rarefaction wave right before the
wave packet. It is formed due to the small discrepancy between
the upstream and downstream values of the Riemann invariant $\la_+$ (see
Fig.~5 and the explanation of the closure conditions in subsection II.C).
It is not difficult to show using standard hydrodynamic reasoning that to
leading order the density jump across this rarefaction wave should be
$\Delta \rho \approx  \delta= (V_m/6)^{3/2}$ (see (\ref{delta})) while
the rarefaction wave speed is calculated as $u+\sqrt{\rho} \approx v+1$.
For the parameters $V_m=0.5$, $v=1$  used in our numerical simulation this
implies $\Delta \rho \approx 0.025$ and at  $t=30$ the predicted position
of the rarefaction wave is about $x=60$. Both predictions completely agree
with the numerical solution.

The agreement between the analytical and numerical solutions seen in  Figs.~13 and 14
is especially remarkable in view of the relatively small width of the potential,
$\sigma=2$, used in the numerical simulations. This width
is comparable with the dispersion length  in the system, which is of order of unity,
so the formal requirement $l =\sigma/2 \gg 1$  of the applicability of the local
hydraulic solution is clearly violated. The robustness of the hydraulic solution
here looks quite surprising and deserves special attention. In this  regard we
note that our analytical construction implies that two potentially conflicting
requirements should be satisfied: the potential barrier should be (i) broad enough
for the hydraulic approximation to be applicable but (ii) not too broad for the
similarity modulation solution could be used for the description of the dispersive shocks (i.e.
the characteristic time of the establishing of the steady transcritical hydraulic
solution should be much less than the characteristic time of the formation of the
dispersive shock). While it might look that these requirements are difficult to
satisfy simultaneously, our numerical simulations show that the resulting analytical
solution works quite well when $\sigma =O(1)$.
We have also performed numerical simulations for $v=1$ with potentials of different
shapes with the conclusion that for potentials with $V_m \lesssim 1$ and
$\sigma = O(1)$ the parameters of the oscillatory structure almost do not
depend on the actual potential width and shape. This implies that for a reasonably
broad range of the barrier potentials the whole wave pattern is characterized only
by the potential strength $V_m$   and the flow velocity $v$, which agrees with the
parametrization in our analytical solution. On the other hand, our simulations
with very broad potentials
$\sigma = O(10)$ show that the  the steady hydraulic transition with constant
jumps outside the potential
does not form within a finite time interval  so the developed quantitative
description of the dispersive
shocks with the aid the similarity  modulation solutions does not apply.

As we see in Figs.~13 and 14, the parameters $v=1$, $V_m=0.5$ correspond to
the case when the downstream dispersive shock is attached to the obstacle. It is
natural to ask at which values of the parameters $v, V_m$ the upstream dispersive shock gets
detached and whether there exists the region of the parameters when both
shocks are detached from the obstacle. To answer these questions,
we first notice that the downstream dispersive shock detaches from the obstacle
when the velocity of the trailing  $(m=0)$ edge of the shock
$v_2(v/2-1, v/2-1, \la_-^d, v/2+1)$ defined by
Eq.~(\ref{13-3}) vanishes.  This condition gives the equation
\begin{equation}\label{A1}
    \tfrac34 v^2-(9+\la_-^d)v-(\la_-^d)^2+6\la_-^d+11=0.
\end{equation}
Taking into account Eq.~(\ref{9-1}) we find $\la_-^d=u^d/2-\sqrt{\rho^d}=u^d-v/2-1$,
and cast this equation to the form
\begin{equation}\label{A2}
    v^2-12v-(u^d)^2+8u^d+4=0 ,
\end{equation}
which gives the value $v^*$ of the ``detachment'' velocity
\begin{equation}\label{A3}
    v^*=6-\sqrt{(u^d-4)^2+16} ,
\end{equation}
where $u^d$ is the greater root of the equation (\ref{9-2}). In the weak potential limit
$V_m\ll1$ we obtain the series expansion
\begin{equation}\label{A4}
    v^*=1+\frac12\sqrt{\frac{3V_m}2}-\frac{V_m}8+\ldots.
\end{equation}
The upstream shock detaches from the obstacle at velocity $v^{**}$ which
satisfies the equation (\ref{15-6}) and again in the weak potential limit
we obtain
\begin{equation}\label{A5}
    v^{**}=1+\frac12\sqrt{\frac{3V_m}2}-\frac{V_m}{24}+\ldots.
\end{equation}
It is easy to see that the region $v^*<v<v^{**}$ is located inside the
transcritical region (\ref{9-6}) and is relatively narrow; its position
is illustrated in Fig.~15.
\begin{figure}[bt]
\begin{center}
\includegraphics[width=8cm,clip]{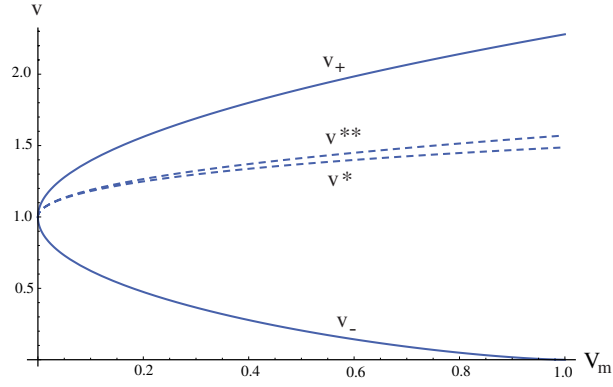}
\caption{Dependence of the velocities $v_-$ and $v_+$ at the
boundaries of the transcritical region (solid lines)
and of the velocities of $v^*$ and $v^{**}$ (dashed lines) on the maximum value $V_m$
of potential. }
\end{center}\label{fig15}
\end{figure}

We have verified the above predictions by constructing numerical solutions of the GP
equation with $V_m=0.5$ and $v=1.4$ ($v^*<v<v^{**}$),  $v=1.8$ ($v>v^{**}$).
The respective plots are presented in Fig.~16.

In Fig.~16a both dispersive shocks are completely developed and actually they both are
detached from the obstacle although because of the small difference between the
soliton edge velocity of the upstream shock and the velocity of the small amplitude
edge of the downstream shock, it takes very long time to reach a well developed hydraulic
transition solution located near $x=0$
solution. Nevertheless we see that both dispersive shocks are not cut off at the edges
close to the obstacle which means their detachment from the obstacle.
In contrast, in Fig.~16b
the downstream dispersive shock is detached from the obstacle whereas the upstream
shock is attached
and is cut off towards the soliton edge. However, one should notice that
the parameters of the dispersive shocks for the velocity $v$ near the upper
boundary $v_+$ of the
transcritical region do not agree well enough with the analytical predictions.
This disagreement has already been
discussed in Section B1 and is due to the violation of our main  supposition that
the flow forms a steady  hydraulic transition over the potential range interval
with the jumps at the both its sides: here we cannot
neglect the time of forming of the hydraulic solution compared with the time of the
development of the dispersive shocks and therefore, the self-similar solutions used in the
analytical theory are not accurate enough. In spite of this reservation,
the developed theory  qualitatively agrees with numerics even in this region: one can easily
distinguish in the numerically obtained pattern all the characteristic ingredients
of our analytical construction, namely, the smooth transition region over the potential
range and downstream and
the upstream dispersive shocks.

\begin{figure}
\begin{center}
\includegraphics[width=8cm,height=5cm,clip]{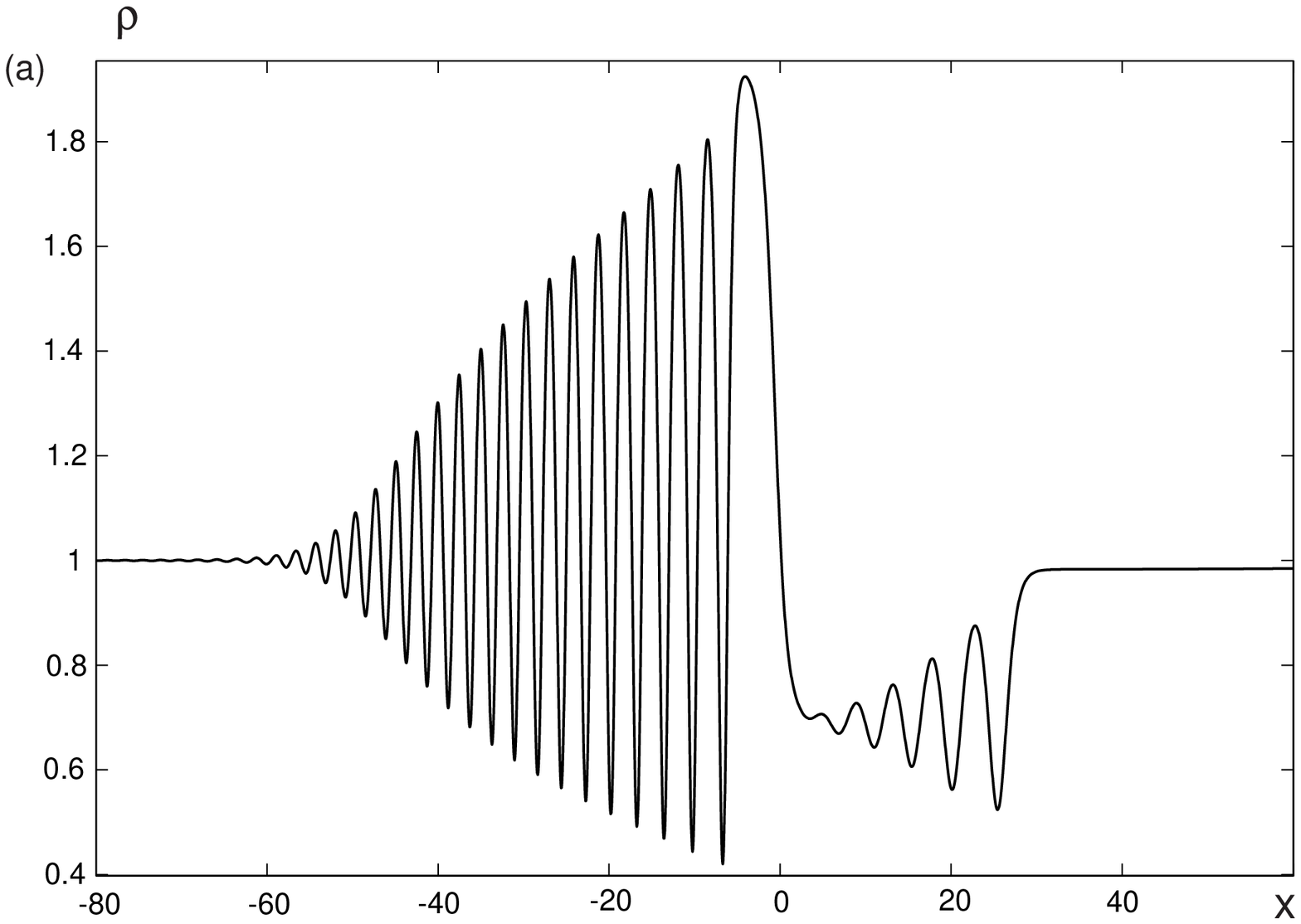}  \qquad
\includegraphics[width=8cm,height=5cm,clip]{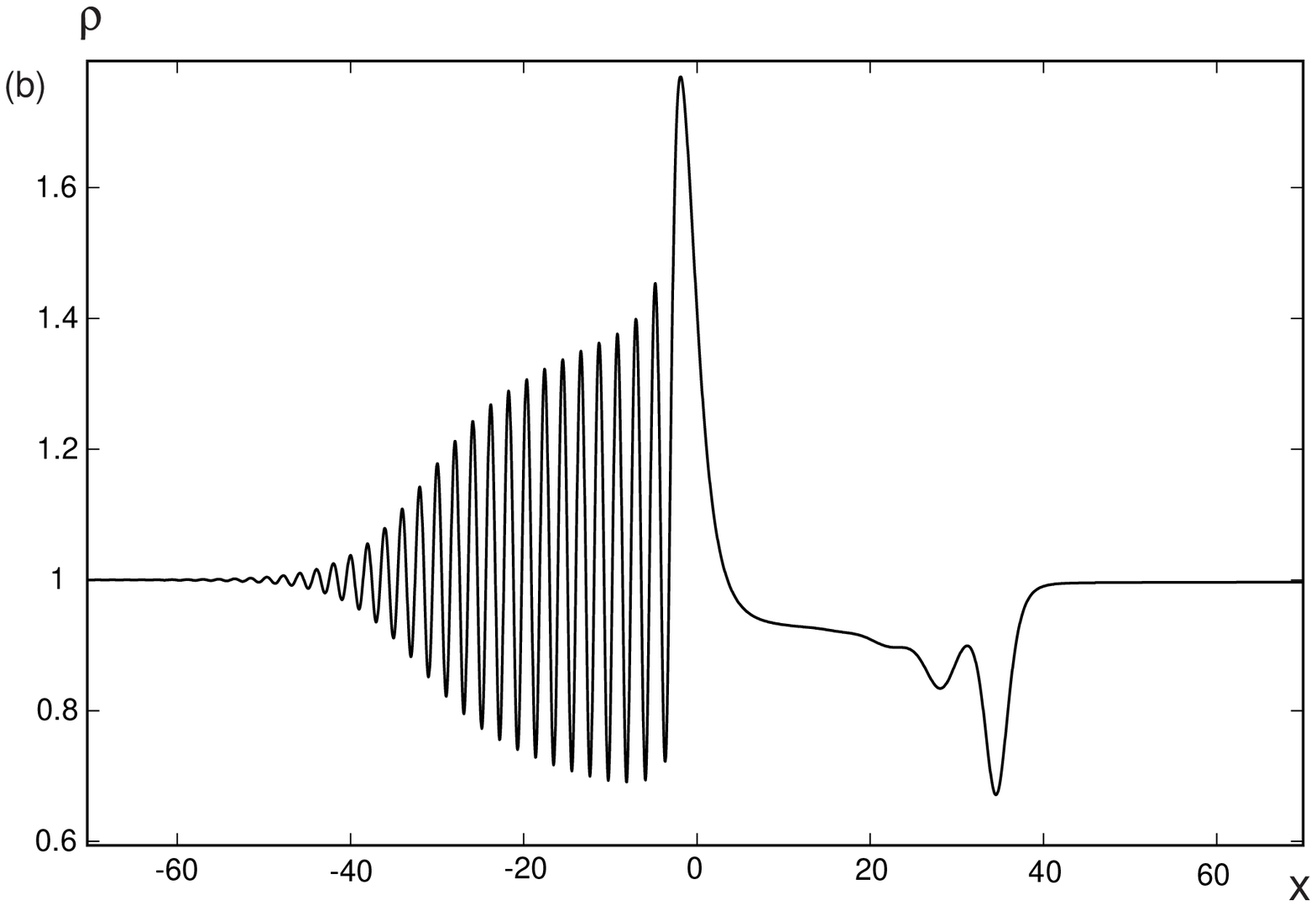}
\caption{Numerical simulation of the condensate density distributions
as functions of $x$ in the wave pattern
generated by the potential barrier (\ref{6-1}) with $V_m=0.5$ and (a) $v=1.4$,
(b) $v=1.8$. Evolution time  $t=45$. }
\end{center}\label{fig16}
\end{figure}

\subsection{Drag force}
We now consider the drag force, i.e. the force exerted on the BEC due to its
motion through the potential barrier. This force can be calculated as
the  spatial mean value of the operator $dV(x)/dx$ over the condensate wave function
(see \cite{pavloff} for a detailed derivation),
\begin{equation}\label{drag}
F_{drag}=\int \limits_{-\infty}^{\infty} \psi^* \frac{\partial V}{\partial x}
\psi dx=\int \limits_{-\infty}^{\infty} \rho \frac{\partial V}{\partial x}dx \, .
\end{equation}
For subcritical and supercritical flows, $\rho=\rho(x)$ is given by the hydraulic
solution (\ref{5-8}) which connects smoothly to $\rho=1$ as $|x| \to \infty$
and satisfies the system (\ref{4-2}). Then integrating the associated Bernoulli equation
\begin{equation}\label{bern}
(\rho u^2 + \rho^2/2)_x + \rho V_x=0
\end{equation}
from $-\infty$ to $+ \infty$, and using that $\rho \to 1$, $u \to v$ as
$|x| \to \infty$ we obtain  $F_{drag}=0$ which is the expression of
BEC superfluidity at sub- and supercritical velocities.
Strictly speaking, the superfluidity is exact for $v<v_-<c_s=1$ where no
excitations are generated. In the supersonic region $v>v_+>c_s=1$ the generation
of excitations exists but it is exponentially small for a slowly varying
obstacles's potential \cite{hakim2} and this generation is neglected in the
hydraulic approximation used here.

\begin{figure}
\begin{center}
\includegraphics[width=7cm,clip]{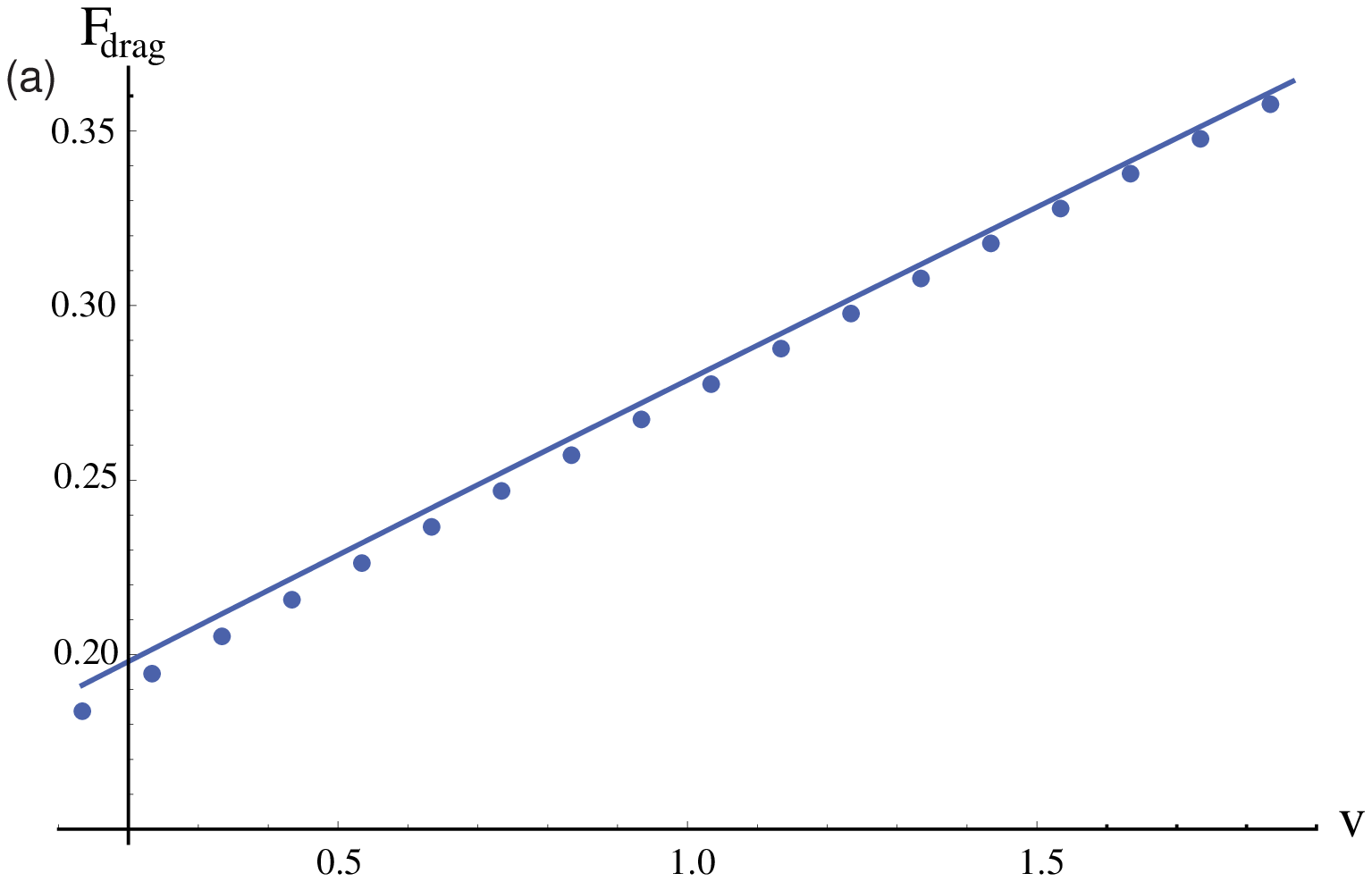}  \qquad
\includegraphics[width=7cm,clip]{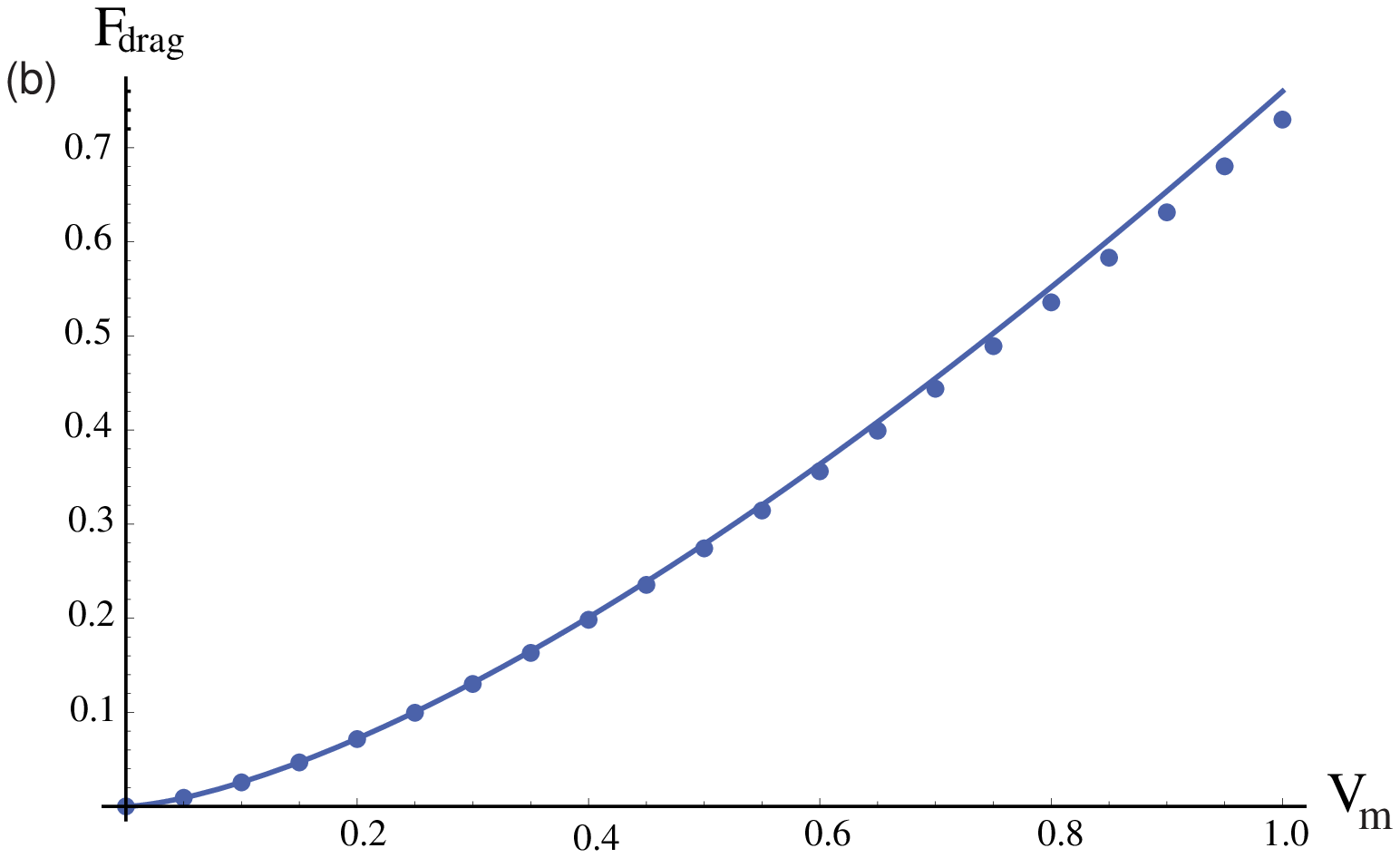}
\caption{ The drag force $F_{drag}$ vs:  a) BEC oncoming flow $v$ in the
transcritical region $(v_-, v_+)$ and
b) potential strength $V_m$ for fixed $v=1$. Solid lines are drawn according
to the approximate Eq.~(\ref{G}) and dots correspond to the full numerical solution of Eq.~(\ref{C}).}
\end{center}\label{fig17}
\end{figure}

For the transcritical regime,  $v_- \le v \le v_+$, there is no global
hydraulic solution  and the integral (\ref{drag}) can be split into three parts
\begin{equation}\label{drag1}
F_{drag}=\int \limits_{-\infty}^{-l} \rho_{d}(x,t) \frac{\partial V}{\partial x} dx
+\int \limits_{-l}^{l} \rho_{tr}(x) \frac{\partial V}{\partial x} dx +
\int \limits_{l}^{\infty} \rho_{u}(x,t) \frac{\partial V}{\partial x} dx.
\end{equation}
Here $\rho_{u,d}(x,t)$ are the unsteady upstream and downstream solutions
describing the density behavior in respective dispersive shocks and
$\rho_{tr}(x)$ is the local transcritical
hydraulic solution (\ref{10-3}) defined on the interval $(-l, l)$.
Assuming that the potential $V(x)$ sufficiently rapidly decays for $|x|>l$
together with its first derivative $V_x$,
one can neglect the contributions of the first and third integrals in (\ref{drag1}).
The remaining second integral can be evaluated again with the use of the
Bernoulli equation (\ref{bern}) to obtain
\begin{equation}\label{A}
F_{drag}=\rho^u(u^u)^2+\tfrac12(\rho^u)^2-\rho^d(u^d)^2-\tfrac12(\rho^d)^2.
\end{equation}
If the downstream or upstream shock is attached to the hydraulic solution, then
the limiting values of the density and the flow velocity in the right-hand
side of Eq.~(\ref{A}) oscillate with time according to the dispersive shock solution
(\ref{11-1}), (\ref{11-4}) considered at  $x=0$ and the parameter $m$ equal to $m^*$
or $m^{**}$ depending on whether the downstream or upstream dispersive shock is attached.
The frequency of the drag force oscillations for the downstream attachment case is given by formula (\ref{f*}) (a similar formula
can be easily obtained for the upstream attachment case as well).
Averaging of the
expression (\ref{A}) over time yields the mean value
of the drag force. The situation simplifies greatly when both dispersive
shocks are detached from the obstacle and $\rho^{u,d}$ and $u^{u,d}$ are
given by the limiting values of the hydraulic solution. Although the corresponding region of the
potential maximum $V_m$ and velocity $v$ values is rather narrow, the discussion
of this case is quite instructive and enables one to estimate the accuracy
of the drag force series expansion in powers of $V_m^{1/2}$ for small $V_m$.

As was shown above, $w=u^{u,d}$ are two roots of the equation (\ref{9-2})
which can be rewritten in a more convenient form after introduction of
new variables
\begin{equation}\label{B}
    w=u_0(1-z),\quad \rho=u_0^2(1+z/2)^2,\quad u_0=(v+2)/3,\quad
    \epsilon=V_m/u_0^2=9V_m/(v+2)^2,
\end{equation}
so that Eq.~(\ref{9-2}) takes the form
\begin{equation}\label{C}
    \tfrac12(1-z)^2+(1+z/2)^2-\tfrac32(1-z)^{2/3}(1+z/2)^{4/3}=\epsilon.
\end{equation}
One can easily derive series expansions of the roots of this equation
in powers of $\epsilon^{1/2}$ to obtain
\begin{equation}\label{D}
    z^u=\sqrt{\frac{2\epsilon}3}-\frac{\epsilon}{18}+\ldots,\quad
    z^d=-\sqrt{\frac{2\epsilon}3}-\frac{\epsilon}{18}+\ldots \, ,
\end{equation}
where the first order terms reproduce actually Eqs.~(\ref{9-4}) after
taking into account inequalities (\ref{9-6}). With the use of
Eq.~(\ref{C}) we represent Eq.~(\ref{A}) as
\begin{equation}\label{E}
    F_{drag}=u_0^4\left[G(z^u)-G(z^d)\right]
\end{equation}
where
\begin{equation}\label{F}
    G(z)=(1-z)^2(1+z/2)^2+\tfrac12(1+z/2)^4.
\end{equation}
Then substitution of (\ref{D}) into (\ref{E}) yields with the
accepted accuracy the expression
\begin{equation}\label{G}
    F_{drag}\cong u_0^4\left(\sqrt{\frac23}\epsilon^{3/2}-\frac5{6\sqrt{6}}\epsilon^{5/2}\right)
    \cong \left(\frac{V_m}6\right)^{3/2}\left[4(v+2)-\frac56 V_m\right],
\end{equation}
where $v$ varies in the transcritical region (\ref{9-6}).
In Fig.~17 we compare the plots of dependence of $F_{drag}$ on $v$ and $V_m$
according to Eq.~(\ref{G}) and calculated by means of the exact numerical solution of
Eq.~(\ref{C}). As we see, the accuracy is very good even for $V_m=1$.

\section{Discussion}

In the experiment \cite{ea07} it was found that the solitons are generated by a
moving potential barrier in the interval of velocities
\begin{equation}\label{16-1}
    0.3\mathrm{mm}/\mathrm{s}<v<0.9\mathrm{mm}/\mathrm{s}.
\end{equation}
This result agrees qualitatively with existence of the finite interval (\ref{6-2})
for which the expanding dispersive shocks are generated. However, we encounter  a
quantitative contradiction if we accept the value of the sound velocity
calculated in \cite{ea07} $c_s=2.1\mathrm{mm}/\mathrm{s}$ as correct,
because in our non-dimensional units the sound velocity is equal to unity and
hence it must be located inside the interval (\ref{6-2}), (see Fig.~2),
or, in dimensional units, inside the interval (\ref{16-1}). This disagreement
can be explained by noticing that the above value of the sound velocity was
calculated in \cite{ea07} according to the expression
\begin{equation}\label{16-2}
    c_s^0=\sqrt{\frac{\rho_0g}{2m}},
\end{equation}
where $\rho_0$ is the condensate density at the center of the trap. Here
\begin{equation}\label{16-3}
    g=4\pi\hbar^2a_s/m
\end{equation}
is the effective coupling constant in the BEC consisting of atoms with
mass $m$ and $s$-wave scattering length $a_s$. But this expression
is correct only for a rarefied enough condensate confined to the
cigar-shaped trap with ``frozen'' radial motion and this condition was
not fulfilled in \cite{ea07}.

Dynamics of a dense BEC is described by the full 3D GP equation, which can
be reduced to some effectively 1D systems much more complicated than
the NLS equation (\ref{3-1}). For example, the variational approach to the
dynamics of the dense BEC was developed in \cite{ks04} where it was shown that  the sound velocity along the axial direction of the trap is given
by the expression (see Eqs.~(67) and (70) in \cite{ks04})
\begin{equation}\label{17-1}
    c_s=c_s^0\cdot\frac{(1+3G/2)^{1/2}}{(1+2G)^{3/4}} ,
\end{equation}
where the parameter $G$ is calculated by the formula
\begin{equation}\label{17-2}
    G=\frac{a_\bot^2}{8\xi^2}\left(\sqrt{1+\left(\frac{a_\bot^2}{8\xi^2}\right)^2}
    +\frac{a_\bot^2}{8\xi^2}\right) .
\end{equation}
Here $a_\bot=\sqrt{\hbar/m\om_\bot}$ is the radial ``oscillator length'' and
$\xi=\hbar/\sqrt{2m\rho_0g}$ is the healing length. Eq.~(\ref{17-1})
reduces to Eq.~(\ref{16-2}) in the limit $G\ll1$ of a rarefied BEC.
In the experiment \cite{ea07} these parameters were equal to
$a_\bot=0.73\cdot 10^{-4}\mathrm{cm}$ and $\xi=0.17\cdot 10^{-4}\mathrm{cm}$,
so that Eq.~(\ref{17-2}) gives $G=11.4$, that is the experiment \cite{ea07}
corresponds to the opposite limit of dense BEC. The sound velocity
calculated according to Eq.~(\ref{17-1}) is equal to $c_s=0.8\mathrm{mm}/\mathrm{s}$,
and this value agrees much better with the interval (\ref{16-1}). Thus, for
quantitative description of the experiment \cite{ea07} the theory of dispersive
shocks in a dense BEC should be developed what is beyond the scope of this paper.
Therefore we shall confine ourselves here to some qualitative remarks only.

The famous Landau criterion \cite{landau} for the loss of  superfluidity was based
on the consideration of linear excitations only and it contradicted to the
experiments with liquid $\mathrm{HeII} $. This discrepancy was explained by
Feynman by taking into the consideration the formation of such nonlinear
structures as vortices in the flow of a superfluid in capillaries or past
obstacles. However, the notion of the threshold velocity below which
the flow is superfluid has not been changed by this modification of the theory.
Taking into account the generation of dispersive shocks in 2D situation \cite{ek06,egk06}
did not change this notion either, since the stationary spatial dispersive shocks
are generated by the supersonic flow of BEC only. The results of the present paper, as well as of the previous papers \cite{hakim,law,hakim2,pavloff,radouani},
show that the situation can be more subtle in the case of 1D flow.
In this case, the flow past a wide barrier leads to the generation of
dispersive shocks in a finite interval of the  flow velocities bounded not only from
below but also from above. Moreover, the lower boundary value of the
velocity $v_-$ can become equal to zero for strong enough barriers, that
is even very slow motions could lead to the generation of solitons. This observation
is in striking contrast with the standard reasonings based on the linear theory
of excitations.

\subsection*{Acknowledgments}

The authors are grateful to Roger Grimshaw for providing an insight into the
transcritical flow past topography problem.  AMK thanks the Royal Society
for the financial support
of his visit to Loughborough University.  YGG and AMK also thank RFBR (grant 09-02-00499)
for partial support.

\end{document}